%% file: BMuMuPaper.tex
\newcommand*{\ATLASLATEXPATH}{latex/}
\documentclass[cernpreprint,texlive=2011,txfonts,UKenglish]{\ATLASLATEXPATH atlasdoc} 

\usepackage[subfigure=true]{\ATLASLATEXPATH atlaspackage}
\usepackage{\ATLASLATEXPATH atlasbiblatex}

\usepackage{\ATLASLATEXPATH atlascontribute}

\usepackage{\ATLASLATEXPATH atlasphysics}

\usepackage{multirow}
\usepackage{wasysym}

\graphicspath{{logos/}{figures/}}

\usepackage{BMuMuPaper-defs}

\input{BMuMuPaper-metadata}
\hypersetup{pdftitle={ATLAS draft},pdfauthor={The ATLAS Collaboration}}

\begin{document}

\maketitle



\input{BMuMuPaper_introduction}
\input{BMuMuPaper_outline}
\input{BMuMuPaper_data}

\input{BMuMuPaper_sel}
\input{BMuMuPaper_bkg}
\input{BMuMuPaper_fakes}

\input{BMuMuPaper_cont}
\input{BMuMuPaper_comp}

\input{BMuMuPaper_bplus}
\input{BMuMuPaper_eff}

\input{BMuMuPaper_sigfit}

\input{BMuMuPaper_br}
\input{BMuMuPaper_conclusions}

\section*{Acknowledgements}

\input{Acknowledgements}



\clearpage
\printbibliography 
\clearpage






\newpage
\input{atlas_authlist}

\end{document}

%% file: BMuMuPaper-metadata.tex
\AtlasTitle{Study of the rare decays of $B^0_s$ and $B^0$ into muon pairs 
from data collected during the LHC Run 1 with the ATLAS detector}

\author{The ATLAS Collaboration}

\AtlasRefCode{BPHY-2012-01}

\PreprintIdNumber{CERN-EP-2016-064}

\AtlasJournal{Eur. Phys. J. C.}

\AtlasAbstract{%
A study of the decays \Bdmumu\ and \Bsmumu\ has been performed using data
corresponding to an integrated luminosity of $25$~fb$^{-1}$ of 
$7$ \TeV\ and $8$ \TeV\ proton--proton collisions collected
with the ATLAS detector during the LHC Run 1. 
For $B^0$, an upper limit on the branching fraction is set at
$\BR(\Bdmumu) < 4.2 \times 10^{-10}$ at $95\%$ confidence level.
For $B^0_s$, the branching fraction
$\BR(\Bsmumu) = \left( 0.9^{+1.1}_{-0.8} \right) \times 10^{-9}$ is measured. 
The results are consistent with the Standard Model expectation with a $p$-value
of $4.8\%$, corresponding to $2.0$ standard deviations. 
}

%% file: BMuMuPaper_introduction.tex
\section{Introduction}
\label{sec:introduction}

Flavour-changing neutral-current (FCNC) processes are highly suppressed 
in the Standard Model (SM), and their study is relevant to
indirect searches for physics beyond the SM.
The branching fractions of the decays \Bmumu\ 
are of particular interest 
because of the additional helicity suppression 
and since they are accurately predicted in the SM:
$\BR(\Bsmumu) = (3.65 \pm 0.23) \times 10^{-9}$ and
$\BR(\Bdmumu) = (1.06 \pm 0.09) \times 10^{-10}$~\cite{Bobeth:2013uxa}.
Significant deviations from these values can arise in models involving 
non-SM heavy particles, such as those predicted in the 
Minimal Supersymmetric Standard Model~\cite{Huang:1998vb,
Hamzaoui:1998nu,Choudhury:1998ze,Babu:1999hn,Choudhury:2005rz}
in extensions such as Minimal Flavour Violation~\cite{D'Ambrosio:2002ex, Buras:2003td},
Two-Higgs-Doublet Models~\cite{Choudhury:2005rz},
and others~\cite{Davidson:2010uu,Guadagnoli:2013mru}.
The CMS and LHCb collaborations have reported the observation of 
\Bsmumu~\cite{Chatrchyan:2013bka,Aaij:2013aka} and evidence of
\Bdmumu, with combined values: 
$\BR(\Bsmumu) = \left(2.8^{+0.7}_{-0.6}\right) \times 10^{-9}$ and 
$\BR(\Bdmumu) = \left(3.9^{+1.6}_{-1.4}\right) \times 10^{-10}$~\cite{CMS:2014xfa}.

This paper reports the result of a search for \Bsmumu\ and \Bdmumu\ 
decays performed using $pp$ collision data corresponding to an integrated
luminosity of 25~fb$^{-1}$, collected at $7$ and $8$~TeV in the full LHC
Run~1 data-taking period using the ATLAS detector.
This analysis supersedes the previous result~\cite{bsmumuv1} based on
2011 data and exploits improved analysis techniques in addition
to the larger dataset.

%% file: BMuMuPaper_outline.tex
\section{Outline} 
\label{sec:outline} 

In order to minimise systematic uncertainties, the \Bsmumu\ and \Bdmumu\ branching
fractions are measured relative to the normalisation decay
$B^+\to J/\psi (\to \mu^+\mu^-)K^+$ that has a known
branching fraction, $\BR (\BpKpJpsi) \times \BR (J/\psi \to \mumu)$.
The procedure needs to be corrected firstly for the ratio of the different probabilities
of the $b$ quark to hadronise with a $u$ or an $s$ quark and this correction
enters as the relative $b$-quark hadronisation probability $f_u/f_d$ or $f_u/f_s$.
Secondly, the different selection efficiencies $\varepsilon_{\mu^+\mu^-}$ and 
$\varepsilon_{J/\psi K^+}$ for the signal and normalisation channels must
be taken into account.
In the simplest form, the \Bdmumu\ (\Bsmumu) branching fraction can be extracted as:
\begin{equation}
\BR (B^0_{(s)}\! \to \!\mu^+ \mu^-) = \frac{N_{d(s)}}{\varepsilon_{\mu^+\mu^-}}
    \times \left[\BR (\BpKpJpsi) \times \BR (J/\psi \to \mumu)\right]
    \frac{\varepsilon_{J/\psi K^+}}{N_{J/\psi K^+}}
    \times \frac{f_{u}}{f_{d(s)}} \nonumber
\end{equation}
where $N_d$ ($N_s$) is the \Bdmumu\ (\Bsmumu) signal yield and $N_{J/\psi K^+}$ is
the \BpKpJpsi\ normalisation yield.
For this study, a modified formula is used to include events selected
in different data-taking periods and with different trigger selections:
\begin{equation}
\BR (B^0_{(s)}\! \to \!\mu^+ \mu^-)
    = N_{d(s)} \times \left[\BR (\BpKpJpsi) \times \BR (J/\psi \to \mumu)\right]
    \times \frac{f_{u}}{f_{d(s)}}
    \times \frac{1}{\mathcal{D}_{\mathrm{norm}}}\,,  \label{eq:BRFormula} \\
\end{equation}
with
\begin{equation}
\mathcal{D}_{\mathrm{norm}} =  \sum_k N^k_{J/\psi K^+} \alpha_k\left(\theInvRho\right)_k \, . 
\label{eq:DnormFormula}
\end{equation}
The denominator $\mathcal{D}_{\mathrm{norm}}$ consists of a sum 
whose index $k$ runs over the data-taking periods and the trigger selections. 
In the sum, the $\alpha_k$ parameter takes into account the different trigger
prescale factors and integrated luminosities in the signal and normalisation channels,
and the ratio of the efficiencies corrects for reconstruction
differences in each data sample $k$.
Signal and reference channel events are selected with similar dimuon triggers
and the normalisation is performed independently for each data-taking period
and trigger category.

The notation used throughout the paper refers to both the stated and
charge-conjugated process, unless otherwise specified. 
The analysis is performed without tagging of the flavour 
$B^0{}\!_{(s)}$ or $\overline{B^0}\!_{(s)}$ at production. 
The yield measurement in the normalisation channel is obtained by summing
$J/\psi K^+$ and $J/\psi K^-$ contributions.

The ATLAS inner tracking system and muon spectrometer are used to reconstruct
and select the event candidates. Details of the detector, trigger, data sets, 
and preliminary selection criteria are discussed in Sections~\ref{sec:data} and~\ref{sec:sel}.

A blind analysis was performed in which data in the dimuon invariant mass region
from $5166$ to $5526$~\MeV\ were removed  
until the procedures
for event selection and the details of signal yield extraction
were completely defined.
The variables used in the event selection, Monte Carlo (MC) tuning and
background studies are discussed in Sections~\ref{sec:data}
to~\ref{sec:bkg}. 
A dedicated strategy was developed to greatly reduce the 
background related to hadrons from $B^0{}\!_{(s)}$ decays which are
misidentified as muons, as described in Section~\ref{sec:fakes}. 
The final sample of candidates is selected using a multivariate classifier,
trained on MC samples, as discussed in Section~\ref{sec:contBDT}.
Checks on the distributions of the variables used in the
multivariate classifier are summarised in Section~\ref{sec:comp}.
Section~\ref{sec:bplus} details the fit procedure to extract the yield of
\BpKpJpsi\ events. As an ancillary measurement to the \BpKpJpsi\ yield
determination, a measurement of the ratio
$\mathcal{B}(B^+ \to J/\psi \pi^+ )/\mathcal{B}(B^+ \to J/\psi K^+)$
is performed, as presented in Subsection~\ref{sec:bpluspikratio}.
The ratio of efficiencies in the signal and the normalisation channels
is presented in Section~\ref{sec:eff}. The extraction of the signal yield is
discussed in Section~\ref{sec:sigfit}, and the branching fractions are
reported in Section~\ref{sec:br}.

%% file: BMuMuPaper_data.tex
\section{ATLAS detector, data and simulation samples}\label{sec:data}
The ATLAS detector\footnote{ATLAS uses a right-handed coordinate system
with its origin at the nominal interaction point. The $z$-axis is along
the beam pipe, the $x$-axis points to the centre of the LHC ring and
the $y$-axis points upward. Cylindrical coordinates $(r,\phi)$ are used
in the transverse plane, $r$ being the distance from the origin and
$\phi$ being the azimuthal angle around the beam pipe. The pseudorapidity
$\eta$ is defined as $\eta = - {\rm ln}[{\rm tan}(\theta/2)]$ where
$\theta$ is the polar angle.}
consists of three main components: an inner detector (ID) tracking system
immersed in a $2$~T axial magnetic field, surrounded by electromagnetic and hadronic
calorimeters and by the muon spectrometer (MS). A full
description can be found in Ref.~\cite{AtlasDet}. 

This analysis is based on the Run~1 data sample recorded in 2011
and 2012 by the ATLAS detector from $pp$ collisions at the LHC at
$\sqrt{s}=7$ and $8$~\TeV, respectively.
Data used in the analysis were recorded during stable LHC beam periods.
Data quality requirements were imposed, notably on the
performance of the MS and ID systems.
The total integrated luminosity of good quality data used in this analysis is
$4.9$~fb$^{-1}$ for the 2011 sample and $20$~fb$^{-1}$ for 2012. 
The average number of reconstructed 
primary vertices (PV) per event, related to multiple proton--proton interactions,
is $6.2$ and $11.4$ in the two years respectively. 

Samples of simulated MC events are used for training and validation
of the multivariate analyses, for the determination of the efficiency
ratios, and for guiding the signal extraction fits. Exclusive MC samples
were produced for the signal channels $\Bsmumu$ and $\Bdmumu$, the
normalisation channel \BpKpJpsi\ (\Jpsimumu),
the \BpPipJpsi\ channel,
and the control channel $\BsJpsiPhi$ ($\Phikk$).
In addition, background studies employ MC samples
of inclusive semileptonic decays $B\to \mu X$,
samples of $\Bs \to K^- \mu^+ \nu$, $B^0 \to \pi^-\mu^+ \nu$, $\Lambda_b \to p \mu^- \overline{\nu}$,
$\Bhh$ decays with $h^{(\prime)}$ being a charged pion or kaon, 
and inclusive decays $B \to J/\psi X$.

Most of the muon pairs in the data sample originate from the uncorrelated
decays of hadrons produced in the hadronisation of a $b$ and a $\bar{b}$ quarks.
To describe this background, defined as combinatorial, a large MC sample 
was generated by selecting specific topologies that dominate it
and producing them semi-exclusively.
The strategy is to consider both the primary decays from $b$ quarks
and the secondary decays from $c$ quarks. Independent samples of
events with forced semileptonic decays or decays 
including muons pairs were generated in all combinations of the two types.
The total number of events in each sample is chosen 
to reproduce the composition of oppositely charged
muon pairs representative of our data.

The MC samples were generated with \Pythia~6~\cite{Sjostrand:2006za}
for studies related to data collected in 2011, and with 
\Pythia~8~\cite{Sjostrand:2007gs} and EvtGen~\cite{EvtGen} for the
2012 sample and the development of multivariate classifiers.
The ATLAS detector and its response are simulated using \Geant~\cite{GEANT4,Aad:2010ah}.
Additional $pp$ interactions in the same and nearby bunch crossings
(pile-up) are included in the simulation.
All simulated samples are reweighted to have the same distribution of
the number of PVs per bunch crossing found in data.

Using the iterative reweighting method described in Ref.~\cite{bsmumuv1},
the simulated samples of the exclusive decays considered are adjusted with
two-dimensional data-driven weights (DDW) to correct for the
differences between MC and data observed in the $\pt^B$ and
and $|\eta^B|$ distributions.
DDW obtained from \BpKpJpsi\ decays are used to correct the
simulation samples in the signal and normalisation channels.
DDW obtained from the \BsJpsiPhi\ control channel are found to
agree with those from \BpKpJpsi\ showing the consistency of the
corrections.

Similarly to the exclusive decays, the large continuum background MC sample
is reweighted via DDW obtained from its comparison with the data in the
sidebands of the signal region.

%% file: BMuMuPaper_sel.tex
\section{Data selection}\label{sec:sel}

For data collected during the LHC Run 1, the ATLAS detector used
a three-level trigger system, consisting of a hardware-based Level-1 trigger,
software-based Level-2 and Event Filter triggers.

A dimuon trigger~\cite{Aad:2012xs, Aad:2014sca} is used to select events.
The 2011 data sample contains events seeded by a Level-1 
dimuon trigger that required a transverse momentum $\pt > 4$~\GeV\ for
both muon candidates. 
Due to the increased pile-up in 2012 data, this dimuon trigger 
was prescaled at the beginning of every fill. 
The effect of prescaling is mitigated by including in the analysis 
events selected by two additional Level-1 triggers scarcely affected by prescaling, 
where tighter selections were applied: 
$\pt > 6$~\GeV\ or $|\eta| < 1.05$ for one of the muons. 
A full track reconstruction of the muon candidates was performed at the 
software trigger levels, where an additional loose selection was applied to the dimuon invariant
mass $m_{\mu\mu}$ and the events were assigned to the $J/\psi$ stream
($2.5 < m_{\mu\mu} < 4.3$~\GeV) or to the $B$ stream ($4.0 < m_{\mu\mu} < 8.5$~\GeV). 

Events from the 2012 dataset are divided into three mutually exclusive
trigger categories and they are assigned to the tightest trigger they passed,
in case they were selected in more than one trigger stream.
We define the following 2012 trigger categories:
\begin{itemize}
    \setlength{\itemsep}{0pt}%
    \setlength{\parskip}{0pt}%
\item[$T_1$:] ``higher threshold'' trigger with \pt $>6$~\GeV\ and $>4$~\GeV\ for
the two muons;
\item[$T_2$:] ``barrel'' trigger with \pt $>4$~\GeV\ for both muon candidates and
  at least one of them 
  with $|\eta| < 1.05$   (and $T_1$ requirement not satisfied);
\item[$T_3$:]  basic dimuon trigger with \pt $>4$~\GeV\ for both muon candidates
  (and $T_1$, $T_2$ requirements not satisfied).
\end{itemize}
Events belonging to a given category are all associated with the same pattern of
Level-1 prescaling. The event sample in the $T_2$ ($T_3$) category has an 
equivalent integrated luminosity equal to $97.7\%$ ($81.3\%$) of the luminosity of the 
$T_1$ category. 
The impact of the trigger Level-1 prescale on the total sample of collected 
events is minor, since the majority of the events belong to the $T_1$ category.

The events in the reference channels \BpKpJpsi \ and $\Bs \to J/\psi \phi$ 
collected in 2012 and used in this analysis belong to a prescaled sample of events,
which was processed together with the signal events.
The effective prescaling factor is equal to $7.3$, and does not affect the sensitivity
of this analysis, given the large number of available events in the normalisation channel.
This factor is included in the $\alpha_k$ parameters in Eq.~(\ref{eq:BRFormula}).

A fourth category is defined for events from the 2011 dataset. They were collected with 
a trigger requirement \pt $>4$~\GeV\ for both muon candidates, and prescaling was 
not applied to this sample.
 
After off-line reconstruction, a preliminary selection is performed on candidates
for $\Bmumu$, $\BpKpJpsi \to \mumu K^+$ and \Bs$ \to J/\psi \phi \to \mumu K^+ K^-$.
In the ID system, muons are required to have at least one hit in the pixel detector, 
five hits in the semiconductor tracker (two hits per each double-sided layer),  
and six hits in the transition-radiation tracker, if $0.1 < |\eta |< 1.9$.
They are also required to be reconstructed in the MS,
and to have $|\eta| < 2.5$ and $\pt>4$~\GeV.
Kaon candidates have to satisfy similar requirements in the ID, except that at least nine instead of six
hits are required in the transition-radiation tracker 
and a looser requirement of $\pt>1$~\GeV\ is imposed.

$B$ meson properties are computed based on a decay vertex fitted to two,
three or four tracks, depending on the decay process to be reconstructed.
The $\chi^2$ per degree of freedom in the vertex fit is required to be
less than six for the $B$ vertex, and less than ten for the $J/\psi \to \mu\mu$ vertex.
The conditions $2915<m(\mu\mu)<3275$~\MeV\ and 
$1005 < m(KK) < 1035$~\MeV\ are required on ID track combinations for the 
$J/\psi \to \mu\mu$ and the $\phi \to K K$ vertices, respectively.
In the \BpKpJpsi\ and \BsJpsiPhi\ fits the reconstructed $J/\psi$ mass is
constrained to the world average value~\cite{pdg14}.

Reconstructed $B$ candidates are required to satisfy $\pt^B>8.0$~\GeV\ 
and $|\eta^B|<2.5$.
The dimuon invariant mass for $\Bmumu$ candidates is calculated using the
combined ID and MS information, in order to improve the mass resolution
in the end-caps with respect to using ID information only~\cite{Aad:2014rra}.

The dimuon mass range considered for the $\Bmumu$ decay is $4766$--$5966$~\MeV\ in which the 
$5166$--$5526$~\MeV\ range is defined as the signal region while 
the low-mass and high-mass regions ($4766$--$5166$ and $5526$--$5966$~\MeV)
are the signal mass sidebands.
For the reference channels, the mass range considered is
$4930$--$5630$ ($5050$--$5650$)~\MeV\ for \BpKpJpsi\ (\BsJpsiPhi)
in which the $5180$--$5380$ ($5297$--$5437$)~\MeV\ range is the peak
region and the two low and high mass ranges are the mass sidebands
used for background subtraction.

The coordinates of the PVs are obtained from charged tracks not used in the decay
vertices, and are transversely constrained to the luminous region of the colliding beams.
The matching of a $B$ candidate to a PV is made by
projecting the direction of the candidate to the point of closest approach
to the collision axis, and choosing the PV with the smallest separation along
$z$. Simulation shows that this method achieves a correct matching probability
of better than $99\%$.

To reduce of the large background in the \Bmumu\ channel
before the final selection based on multivariate classifiers, a loose collinearity
requirement is applied between the momentum of the $B$ candidate
($\vec{p}^B$) and the spatial separation 
between the PV and the decay vertex ($\overrightarrow{\Delta x}$). 
The absolute value of the difference in azimuthal angle $\alpha_{2\mathrm{D}}$ is required to be smaller than 
1.0 rad.  Using the difference in rapidity $\Delta\eta$, the combination 
$\Delta R = \sqrt{\alpha_{2\mathrm{D}}{}^{2} + \Delta\eta^{2}}$ is required to be smaller
than $1.5$.
These requirements reduce the background by a factor of $0.4$, with a signal
efficiency of $95\%$.

After the preliminary selection, approximately $2.6\times 10^6$ $\Bmumu$
and $2.3 \times10^6$ $\BpKpJpsi$ candidates are found in the signal regions.

%% file: BMuMuPaper_bkg.tex
\section{Background composition}
\label{sec:bkg}

The background to the \Bmumu\ signal can be separated into three categories:
\begin{description}
    \setlength{\itemsep}{0pt}%
    \setlength{\parskip}{0pt}%
\item[] {\em combinatorial} background, the dominant component due to pairs of uncorrelated
muons and characterised by a small dependence on the dimuon invariant mass;
\item[] {\em partially reconstructed}  $B\to\mu\mu X$ decays, characterised by non-reconstructed
final-state particles ($X$) and thus accumulating in the low-mass sideband;
\item[] { \em peaking} background, due to $B^0{}\!_{(s)}\to h\,h^\prime$ decays, with both hadrons misidentified as muons.
\end{description}

The combinatorial background consists mainly of
muons independently produced in the fragmentation and decay trees of a  
$b$ and a $\overline{b}$ quark ({\em opposite-side} muons). 
It is studied in the signal mass sidebands, and it is found to be
correctly described by the inclusive MC sample of combinatorial background from
semileptonic decays of $b$ and $c$ hadrons. Section~\ref{sec:comp} contains
data--MC comparisons for the combinatorial background.
As discussed in Section~\ref{sec:contBDT}, a multivariate classifier
trained on MC samples is used to reduce this component. 

The partially reconstructed decays consist of several topologies:
(a) {\em same-side} (SS) combinatorial background from decay cascades
($b \to c \mu\nu \to  s(d) \mu\mu\nu\nu$);  
(b) {\em same-vertex} (SV) background from $B$ decays containing a muon pair
(e.g. $B^0 \to K^{*0}\mu\mu$, $B \to J/\psi X \to \mu\mu\mu X^\prime$);
(c) $B_c$ decays (e.g. $B_c \to J/\psi \mu \nu \to \mu\mu\mu\nu$);
(d) semileptonic $b$-hadron decays where a final-state hadron is
misidentified as a muon.

Inclusive MC samples of SS events, SV events, and $B_c \to J/\psi \mu \nu$ decays
were generated together with the large sample of combinatorial background.
All subsamples have a dimuon mass distribution accumulating below the mass range
considered in this analysis. The high-mass tail extends to the signal region
and becomes a significant fraction of the background only after applying a selection
against the combinatorial background.

\begin{figure}[!htb]
  \begin{center}
    \subfigure[]{
      \hspace*{-0.5cm}
      \includegraphics[width=0.52\textwidth]{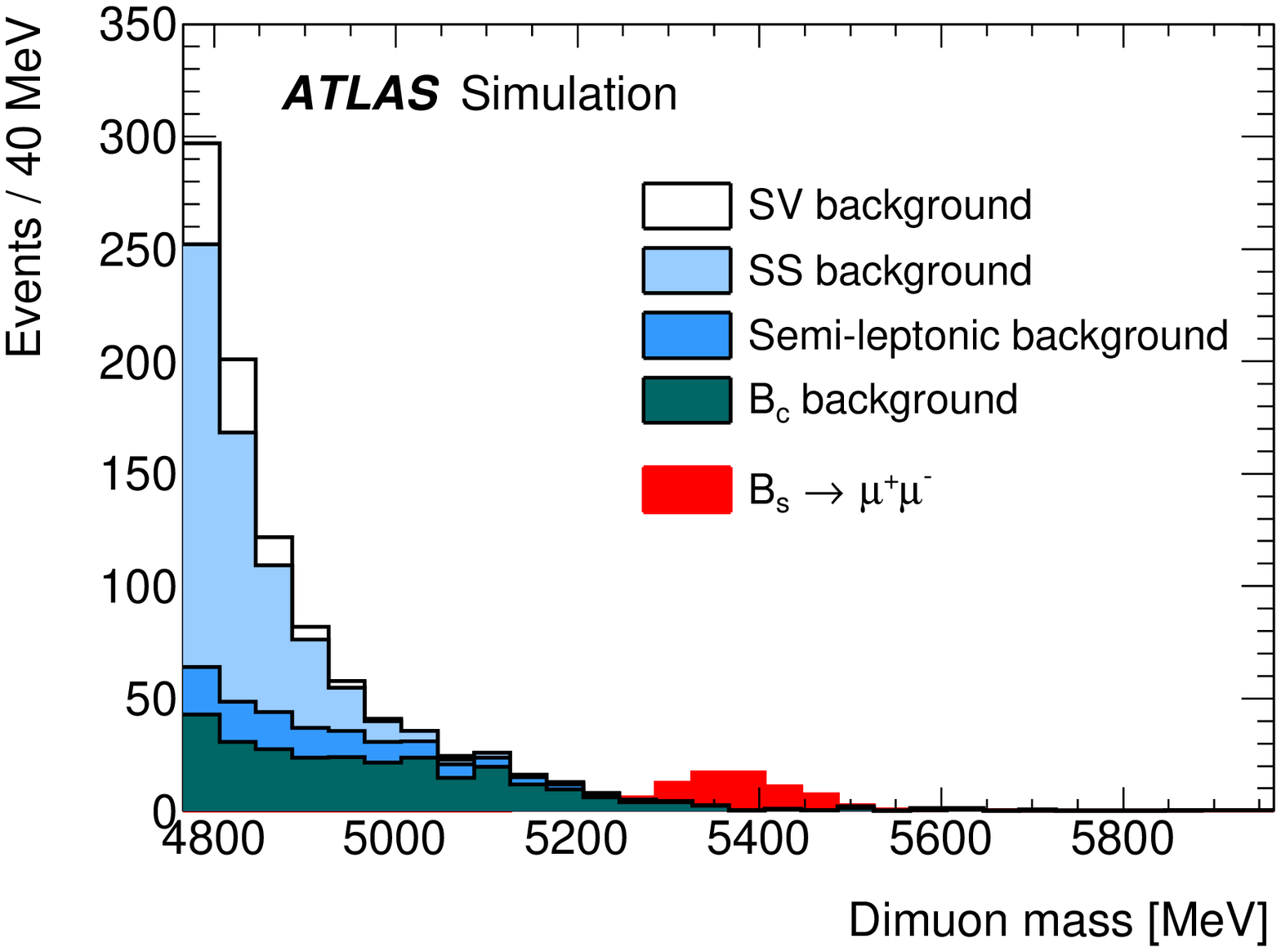}
    }
    \subfigure[]{
      \hspace*{-0.5cm}
      \includegraphics[width=0.52\textwidth]{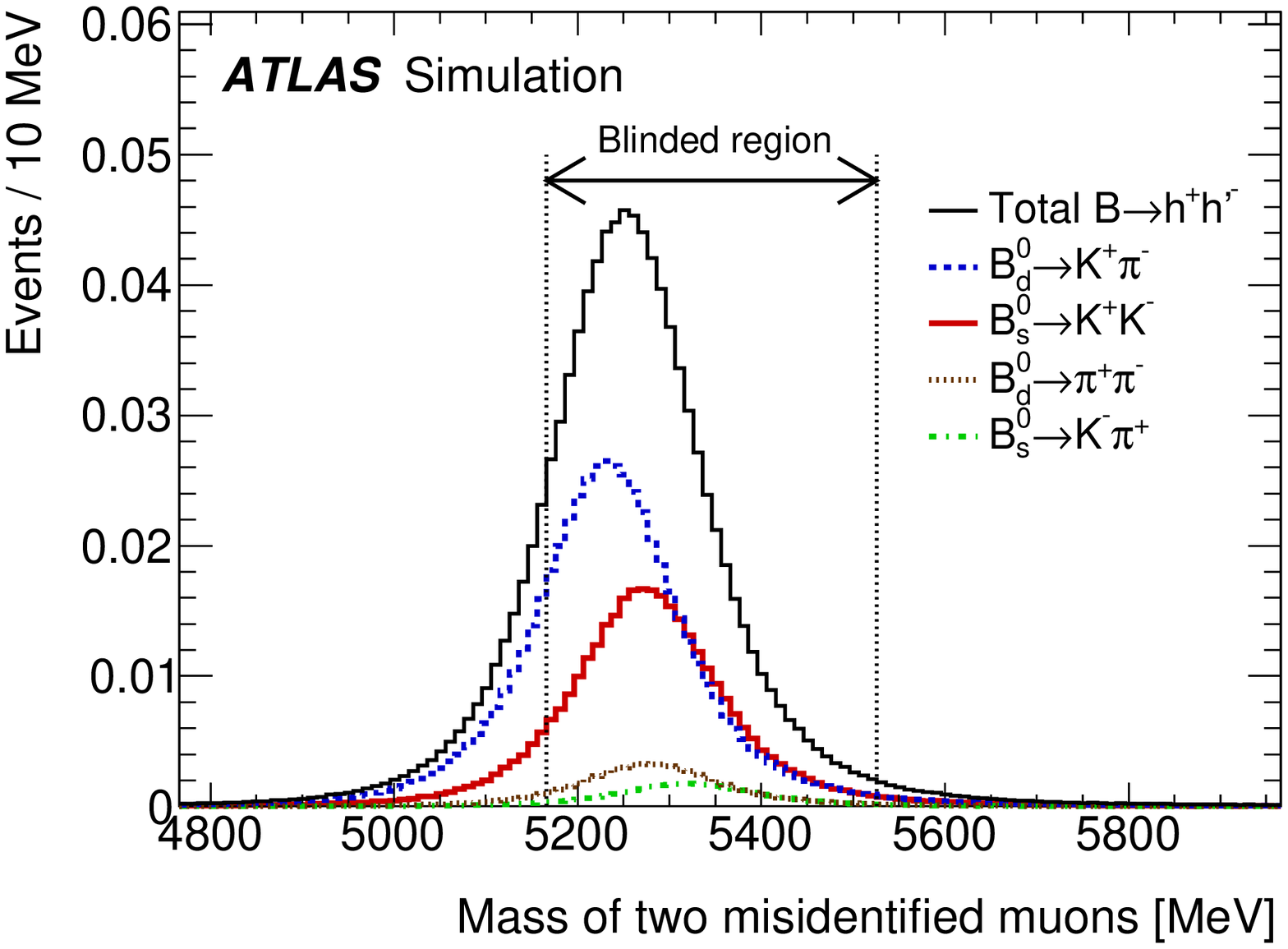}
      \hspace*{-0.6cm}
    }
    \caption{(a) Dimuon mass distribution for the partially reconstructed
background, from simulation, before the final selection against continuum is
applied but after all other requirements. 
The different components are shown as stacked histograms, normalised according
to world-averaged measured branching fractions. The SM expectation for the
\Bsmumu\ signal is also shown for comparison (non-stacked).
Continuum background is not included here.
(b) Invariant mass distribution of the peaking background components \Bhh,
after the complete signal selection is applied. 
In both plots the distributions are normalised to the expected yield for the 
integrated luminosity of 25 fb$^{-1}$.}
\label{fig:SS-SV}
\end{center}
\end{figure}
The semileptonic decays with final-state hadrons misidentified as muons
consist mainly of three-body charmless decays $B^0 \to \pi\mu\nu$,
$B^0_s \to K\mu\nu$ and $\Lambda_b \to p\mu\nu$ in which the tail
of the mass distribution extends to the signal region.
Due to branching fractions of the order of $10^{-6}$,
this background is not large, and is further reduced by the dedicated muon identification
requirements, discussed in Section~\ref{sec:fakes}.
The MC mass distributions of these partially reconstructed decay topologies are shown in
Figure~\ref{fig:SS-SV}(a) after applying the preliminary selection criteria
described in Section~\ref{sec:sel}.

Finally, the peaking background is due to $B^0{}\!_{(s)}$ decays containing 
two hadrons misidentified as muons, which populate the signal region 
and cannot be observed in the sidebands. This is shown in
Figure~\ref{fig:SS-SV}(b).

%% file: BMuMuPaper_fakes.tex
\section{Multivariate analysis for reducing hadron misidentification}
\label{sec:fakes}

In the preliminary selection, muon candidates are formed from the combination 
of tracks reconstructed independently in the ID and MS~\cite{Aad:2009wy}.
The performance of the muon reconstruction in ATLAS is presented in Ref.~\cite{Aad:2014rra}.  
Additional studies were performed for this analysis to minimise and
evaluate the amount of background related to hadrons erroneously identified as muons. 

Detailed simulation studies were performed for the channels \Bhh\ 
and $\Lambda_b \to ph$, with $h^{(\prime)} = \pi^\pm , K^\pm$. 
A full \Geant-based simulation~\cite{GEANT4} in all systems of the ATLAS detector
is used for this purpose.
The vast majority of background events from particle misidentification are due to decays
in flight of kaons and pions, in which the muon receives most of the energy of the meson.
Hence, despite the notation of {\em fake} muons, this background is generally related
to true muons measured in the MS, but not produced promptly in the decay of a $B$ meson.
The contribution from hadronic punch-through into the MS is expected from simulation to
amount only to 3\% (8\%) of the total number of fake candidates from kaons (pions).

The simulation shows that after the preliminary selection the probability for a kaon
(pion) to be misidentified as a muon is 0.4\% (0.2\%).
This fraction is found to be largely independent of the
transverse momentum and rapidity of the track, as well as other variables related
to the underlying event or pile-up. The misidentification rate for protons 
is found to be negligible ($<$ 0.01\%).

The muon candidate is further required to match the trigger requirements, resulting
in a reduction in the number of retained tracks by a factor 0.58, and to pass an
additional multivariate selection, implemented as a boosted decision tree
(BDT)~\cite{TmvaManual}.
This selection, referred to as {\em fake-BDT}, is based on variables described in 
Table~\ref{tab:fake-BDT} and it is built and trained on the MC samples mentioned above.
The BDT training is done using a multivariate analysis tool (TMVA)~\cite{TmvaManual}.
The fake-BDT selection is tuned for a 95\% efficiency for muons in the signal sample, 
and achieves an average reduction of the hadron misidentification by a factor 0.37, 
determined with independent MC samples.
The resulting final value of the misidentification probability is equal to 0.09\% for
kaons and 0.04\% for pions. 

\begin{table}[!bt]
\small
  \caption{Description of the eight variables used in the discrimination between signal
  muons and those from hadron decays in flight and punch-throughs.}
  \begin{center}
  \begin{tabular}{p{0.2cm} p{10cm}}
    \toprule \\[-9pt]
    1. & Absolute value of the track rapidity measured in the ID.\\[5pt]
    2. &Ratio $q/p$ (charge over momentum) measured in the MS.\\[5pt]
    3. & Scattering curvature significance: maximum variation of the track
    curvature between adjacent layers of the ID.\\[5pt]
    4. &$\chi^2$ of the track reconstruction in the MS.\\[5 pt]
    5. &Number of hits used to reconstruct the track in the MS.\\[5pt]
    6. &Ratio of the values of $q/p$ measured in the ID and
    in the MS, corrected for the average energy loss in the calorimeter.\\[5pt]
    7. & $\chi^2$ of the match between the tracks reconstructed in the ID and MS.\\[5pt]
    8. &Energy deposited in the calorimeters along the muon trajectory
    obtained by combining ID and MS tracks.\\[5pt]
    \bottomrule
  \end{tabular}
  \label{tab:fake-BDT}
\end{center}
\end{table}

The background due to \Bhh, with double misidentification $h h^\prime \to \mu\mu$,
has a distribution in the reconstructed mass peaking at 5250~\MeV, close to the
\Bs\ mass and is effectively indistinguishable from the \Bz\ signal 
(see Figure~\ref{fig:SS-SV}(b)).
Beyond the muon and fake-BDT selection, 
these events have the same acceptance and
selection efficiency as the \Bmumu\  signal. Therefore, the expected number of
peaking-background events can be estimated from the number of observed
\BpKpJpsi\ events, in a way analogous to what is done for the signal,
using Eq.~(\ref{eq:BRFormula}).
World average~\cite{pdg14} values for the branching fractions of \Bz\ and \Bs\ into
$K\pi$, $KK$ and $\pi\pi$ are used, together with the hadron
misidentification probabilities obtained from simulation.  The resulting total
expected number of peaking-background events, after the final selection
(including a multivariate cut against $\mu^+\mu^-$ continuum background, 
the {\em continuum}-BDT 
discussed in Section~\ref{sec:contBDT}), 
is equal to 0.7,  with a 10\% uncertainty from the normalisation procedure. 

The simulation of hadron misidentification was validated and calibrated with studies
performed on data. The fractions of fake muons after the preliminary selection were
evaluated on samples of $\phi \to K^+K^-$ and \BpKpJpsi\ events, and found to be
consistent with the simulation within a factor $1.2\pm0.2$. 
This factor and its square $1.4 \pm 0.5$ are used as scale correction
and systematic uncertainty in the single and double misidentification
probability, respectively. Hence, the expected number of peaking background
events is equal to  $1.0\pm 0.4$.

A further test of the peaking background was performed on the 
final sample of \Bmumu\ candidates.  Inverting the selection applied with the 
fake-BDT, the number of events containing real muons is largely reduced,
while the number of peaking-background events is approximately three times
larger than in the sample obtained with the nominal selection. 
The result of a fit to the background-enhanced sample 
is $0.5 \pm 3.0$ events of peaking background, in good agreement with the expectation. 

The efficiency of the fake-BDT selection when applied to muons from \Bmumu\ decays 
was tested on the sample of \BpKpJpsi\ candidates selected in data. 
The value from MC simulation was found to be accurate to better than $1\%$.

Besides the peaking background, the selection with the fake-BDT also reduces the
semileptonic contributions with a single misidentified hadron. The expected
number of events from $\Bz \to \pi\mu\nu$ and $\Bs \to K\mu\nu$ in the final
sample is $107 \pm 27$. The $\Lambda_b \to p\mu\nu$ contribution is negligible
due to the smaller production cross section and the fake rejection for protons at the
level of $10^{-5}$.

%% file: BMuMuPaper_cont.tex
\section{Multivariate analysis for continuum discrimination}
\label{sec:contBDT}

A multivariate analysis, implemented as a BDT, is employed
to enhance the signal relative to the combinatorial background.
This classifier, referred to as the {\em continuum-BDT}, 
is based on the $15$ variables described in Table~\ref{tab:15vars}.  
The discriminating variables can be classified into three groups:
(a) $B$ meson variables, related to the reconstruction of the decay vertex and 
to the collinearity between $\overrightarrow{p}^B$ and the separation between
production and decay vertices $\Delta\overrightarrow{x}$;
(b) variables describing the muons forming the $B$ meson candidate;
and (c) variables related to the rest of the event.
The selection of the variables aims to optimise the discrimination power
of the classifier, while minimising the dependence on the invariant mass
of the muon pair. 

Most of the discriminating variables are part of the set used in the previous
analysis based on data collected in 2011~\cite{bsmumuv1}, while others
were modified or added, exploiting the statistical power of the large
samples of MC events used for training and validating the classifier. 
To minimise the dependence of the classifier on the effects of the pile-up,
requirements of compatibility with the same vertex matched to the 
dimuon candidate are placed on the additional tracks considered for the variables 
$I_{0.7}$, DOCA$_{\mathrm{xtrk}}$ and $N^{\mathrm{close}}_{\mathrm{xtrk}}$ .

The correlation between the discriminating variables 
was studied in the MC samples for signal and combinatorial background 
discussed in Section~\ref{sec:data},
 and on data from the sidebands of the $\mu^+\mu^-$ mass distribution.
Different degrees of correlation are 
present, with significant linear correlation among the variables 
$\chi^{2}_{{\mathrm{PV,DV}}\,xy}$, $L_{xy}$, $|d_{0}|^{\mathrm{max}}$-sig.,
$|d_{0}|^{\mathrm{min}}$-sig. and $\chi^2_{\mu,\mathrm{xPV}}$.
Conversely, the variables 
IP$_B^{3\mathrm{D}}$, DOCA$_{\mu\mu}$ and $I_{0.7}$ have negligible correlation
with any of the others used in the classifier. 

\begin{table}[!tb]
\small
\begin{center}
\caption{Description of the $15$ variables used in the discrimination between signal
and combinatorial background. When the BDT classifier is applied to \BpKpJpsi\ and
\BsJpsiPhi\ candidates, the variables related to the decay products of the $B$
mesons refer only to the muons from the decay of the $J/\psi$.}
\input{BMuMuPaper_BDTtable.tex}
\label{tab:15vars}
\end{center}
\end{table}

The MC sample for signal and the large MC sample of semileptonic decays of hadrons containing
$b$ or $c$ quarks are used for training and testing the classifier.
As discussed in Section~\ref{sec:data}, signal and background samples are reweighted according to the 
distributions of \pt and $|\eta|$ of the dimuon  
and of the number of reconstructed PVs observed in data.
To correctly reproduce the data collected in 2012, the MC events belonging to different trigger streams 
are reweighted according to the relative equivalent luminosity
and to two different versions of the Level-2 muon reconstruction algorithm used during the data taking.
The BDT training is done using TMVA~\cite{TmvaManual}.

\begin{figure}[!htb]
  \begin{center}
    \includegraphics[width=0.54\textwidth]{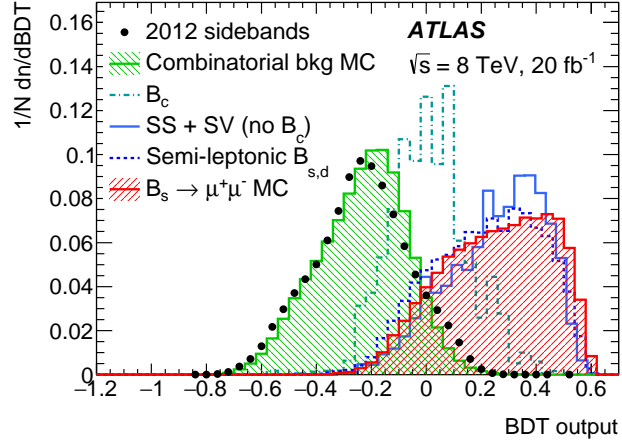} 
    \caption{
      Continuum-BDT distribution for the signal and background
      events: signal $B^0{}\!_{(s)}$, partially reconstructed $B$ events (SS+SV),
      \Bc\ decays and continuum. The solid histograms are obtained from MC simulation,
      while the points represent data collected in the sidebands.
      All distributions are normalised to unity.
      The distributions are shown after the preliminary selection, and before
      applying any reweighting to the variables used in the classifier.  
    }
    \label{fig:contBDT}
  \end{center}
\end{figure}

Figure~\ref{fig:contBDT} shows the distribution of the BDT output variable 
for signal and background, separately for combinatorial background and partially reconstructed events. 
Also shown is the BDT distribution for dimuon candidates from data, from the sidebands of the
mass distribution.  
With the normalisation observed in the data before the application of the BDT selection, 
the combinatorial background dominates over the other sources by a factor 
of about 2000 in the number of events contained in the full mass window and full range of the continuum-BDT output.
In both the signal and background MC samples, the absolute value of the linear correlation
coefficient between the BDT output and the dimuon invariant mass is smaller than 1\%.  

The final selection requires a continuum-BDT output value larger than $0.24$, corresponding
to a signal relative efficiency of $54\%$ (see Section~\ref{sec:sigfit}), and
to a reduction of the combinatorial background by a factor of about $10^{-3}$.

%% file: BMuMuPaper_BDTtable.tex
\begin{tabular}{p{2cm} p{13cm}}
  \toprule \\[-9pt]
  Variable  &  Description \\[2pt] 
  \midrule \\[-9pt]
  $\pt^{B}$ & Magnitude of the $B$ candidate transverse momentum $\overrightarrow{\pt}^B$.\\[5pt]
   $\chi^{2}_{{\mathrm{PV,DV}}\;xy}$ & Significance of the separation 
  $\overrightarrow{\Delta x}$ between production ({\em i.e.} associated PV) and decay (DV)
  vertices in the transverse projection:
   $\overrightarrow{\Delta x}_\text{T}\!\cdot \! \Sigma_{\overrightarrow{\Delta x}_\text{T}}^{\phantom{x}-1}\!\cdot\!\overrightarrow{\Delta x}_\text{T}$,
  where $\Sigma_{\overrightarrow{\Delta x}_\text{T}}$ is the covariance matrix. 
  \\[5pt]
  $\Delta{R}$ & three-dimensional opening between $\overrightarrow{p}^{B}$ and 
  $\overrightarrow{\Delta x}$: $\sqrt{\alpha_{2\mathrm{D}}{}^{2} + \Delta\eta^{2}}$ 
  \\[3pt]
  $|\alpha_{2\mathrm{D}}|$ & Absolute value of the angle between $\overrightarrow{\pt}^{B}$ and
  $\overrightarrow{\Delta x}_\text{T}$ (transverse projection). 
  \\[3pt]      
  $L_{xy}$ & Projection of $\overrightarrow{\Delta x}_\text{T}$ along the direction of $\overrightarrow{p}_\text{T}^{B}$:
  $(\overrightarrow{\Delta x}_\text{T}\!\cdot\!\overrightarrow{\pt}^{B})/|\overrightarrow{\pt}^{B}|$. \\[3pt]
  IP$_B^{3\mathrm{D}}$ & three-dimensional impact parameter 
  of the $B$ candidate to the associated PV.\\[3pt]
  DOCA$_{\mu\mu}$ & Distance of closest approach (DOCA) of the two tracks forming the $B$ candidate (three-di\-men\-sion\-al).
  \\[3pt]
  $\Delta\phi_{\mu\mu}$ & Difference in azimuthal angle between the momenta of the two tracks forming the $B$ candidate.
  \\[3pt]        
  $|d_{0}|^{\mathrm{max}}$-sig. & Significance of the larger absolute value of the impact
  parameters to the PV of the tracks forming the $B$ candidate,  in the transverse plane.
  \\[3pt]
  $|d_{0}|^{\mathrm{min}}$-sig. & Significance of the smaller absolute value of the impact
  parameters to the PV of the tracks forming the $B$ candidate, in the transverse plane.
  \\[3pt]
  $P_{\text{L}}^{\mathrm{min}}$ & Value of the smaller projection of the momenta of the muon candidates
  along $\overrightarrow{\pt}^B$. 
  \\[3pt]
  $I_{0.7}$ 
  & Isolation variable defined as ratio of $|\overrightarrow{\pt}^{B}|$ to the sum of $|\overrightarrow{\pt}^{B}|$ and of 
  the transverse momenta of all additional tracks contained 
  within a cone of size $\Delta{R} < 0.7$ around the $B$ direction.
  Only tracks with $\pt > 0.5$~\GeV\ and matched to the same
  PV as the $B$ candidate are included in the sum.
  \\[3pt]
  DOCA$_{\mathrm{xtrk}}$ & DOCA of the closest additional track to the decay vertex of the $B$ candidate.
  Tracks matched to a PV different from the $B$ candidate are excluded. \\[3pt]
  $N^{\mathrm{close}}_{\mathrm{xtrk}}$ & Number of additional tracks compatible with the decay vertex (DV) of the 
  $B$ candidate  with $\ln(\chi^2_{\mathrm{xtrk,DV}})\! < \! 1$. 
  The tracks matched to a PV different from the $B$ candidate are excluded.
  \\[3pt]
  $\chi^2_{\mu,\mathrm{xPV}}$ & Minimum $\chi^2$ for the compatibility of a muon in the $B$ candidate 
  with a PV different from the one associated with the $B$ candidate.\\[5pt] 
\bottomrule
\end{tabular}

%% file: BMuMuPaper_comp.tex
\section{Data--MC comparisons}
\label{sec:comp}

\begin{figure}[!htb]
  \begin{center}
    \subfigure[]{
      \hspace*{-0.5cm}
      \includegraphics[width=0.52\textwidth]{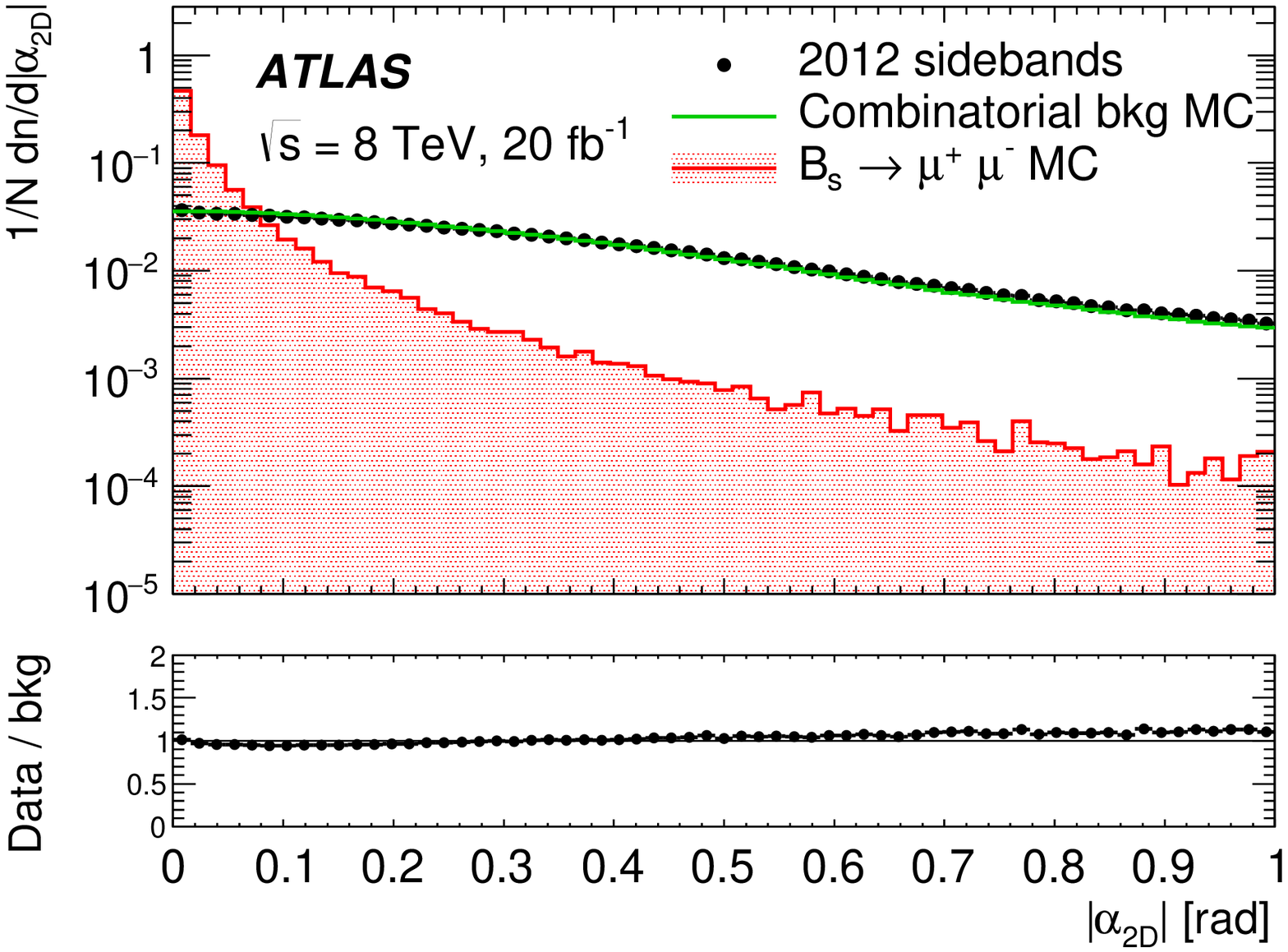}
    }
    \subfigure[]{
      \hspace*{-0.5cm}
      \includegraphics[width=0.52\textwidth]{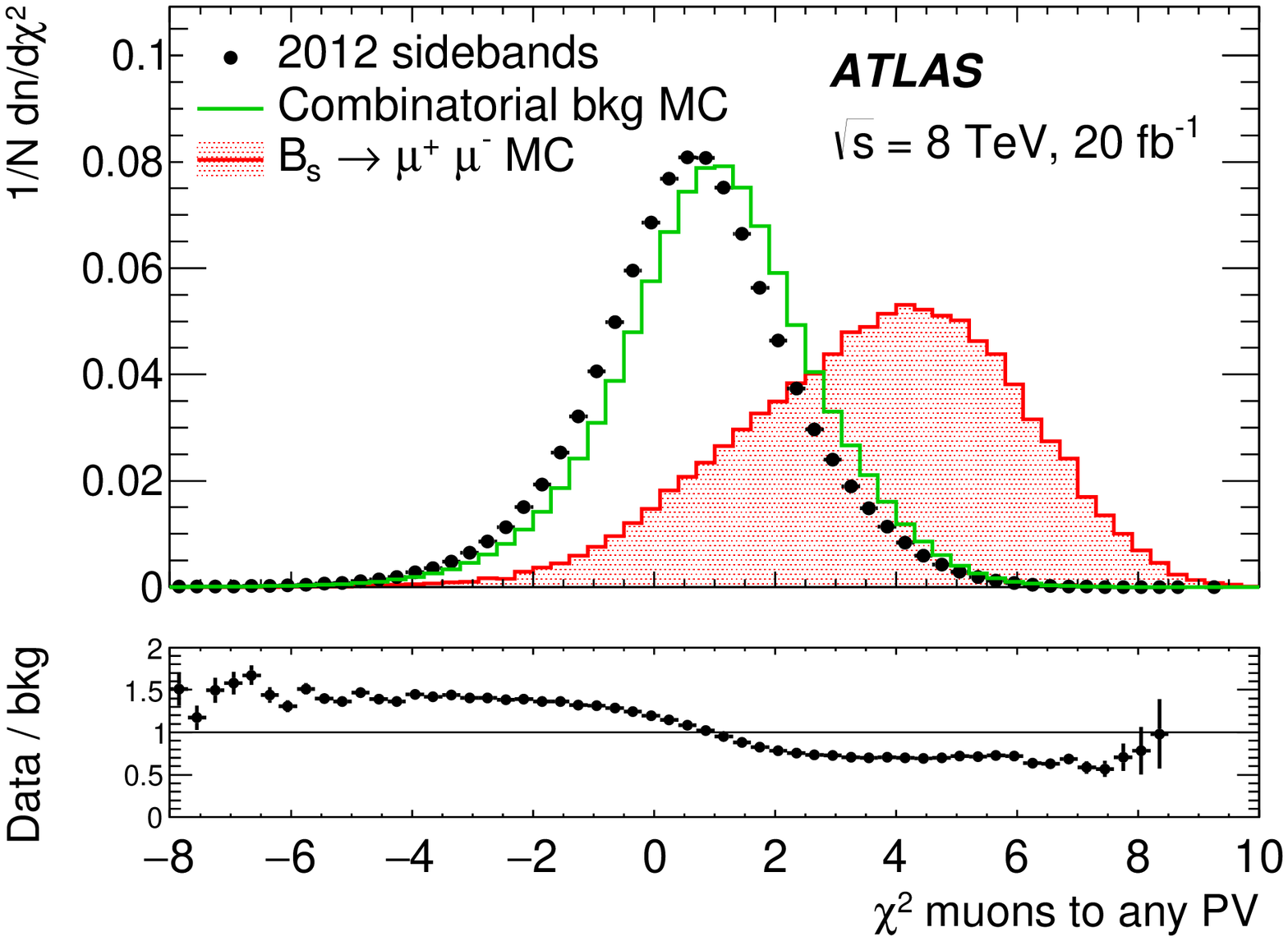}
      \hspace*{-0.6cm}
    }
\caption{Data and continuum MC distributions of $|\alpha_{2\mathrm{D}}|$ (a)
  and $\chi^2{}_{\mu,\mathrm{xPV}}$ (b) variables (see Table~\ref{tab:15vars}).
  The dots correspond to the 2012 sideband data, while 
  the continuous-line histogram corresponds to 
  the continuum MC distribution, normalised to the number of data events. 
  The filled-area histogram shows the signal MC distribution
  for comparison. Discrepancies between MC events and sideband data like the one 
  observed for $\chi^2_{\mu,\mathrm{xPV}}$ do not compromise significantly 
  the optimisation of the continuum-BDT classifier.}
\label{fig:maincompcont}
\end{center}
\end{figure}

The distributions of the discriminating variables are used to compare the MC
sample of semileptonic decays with data in the dimuon sidebands.
Figure~\ref{fig:maincompcont} shows the distributions for two discriminating
variables. Agreement with the sideband data is fair and the discrepancies
observed do not compromise the use of this MC background sample 
for the purpose of training the continuum-BDT. 
The continuum MC simulation is not used for computation of efficiencies
or normalisation purposes.

\begin{figure}[!h]
  \begin{center}
    \subfigure[]{
      \hspace*{-0.5cm}
      \includegraphics[width=0.52\textwidth]{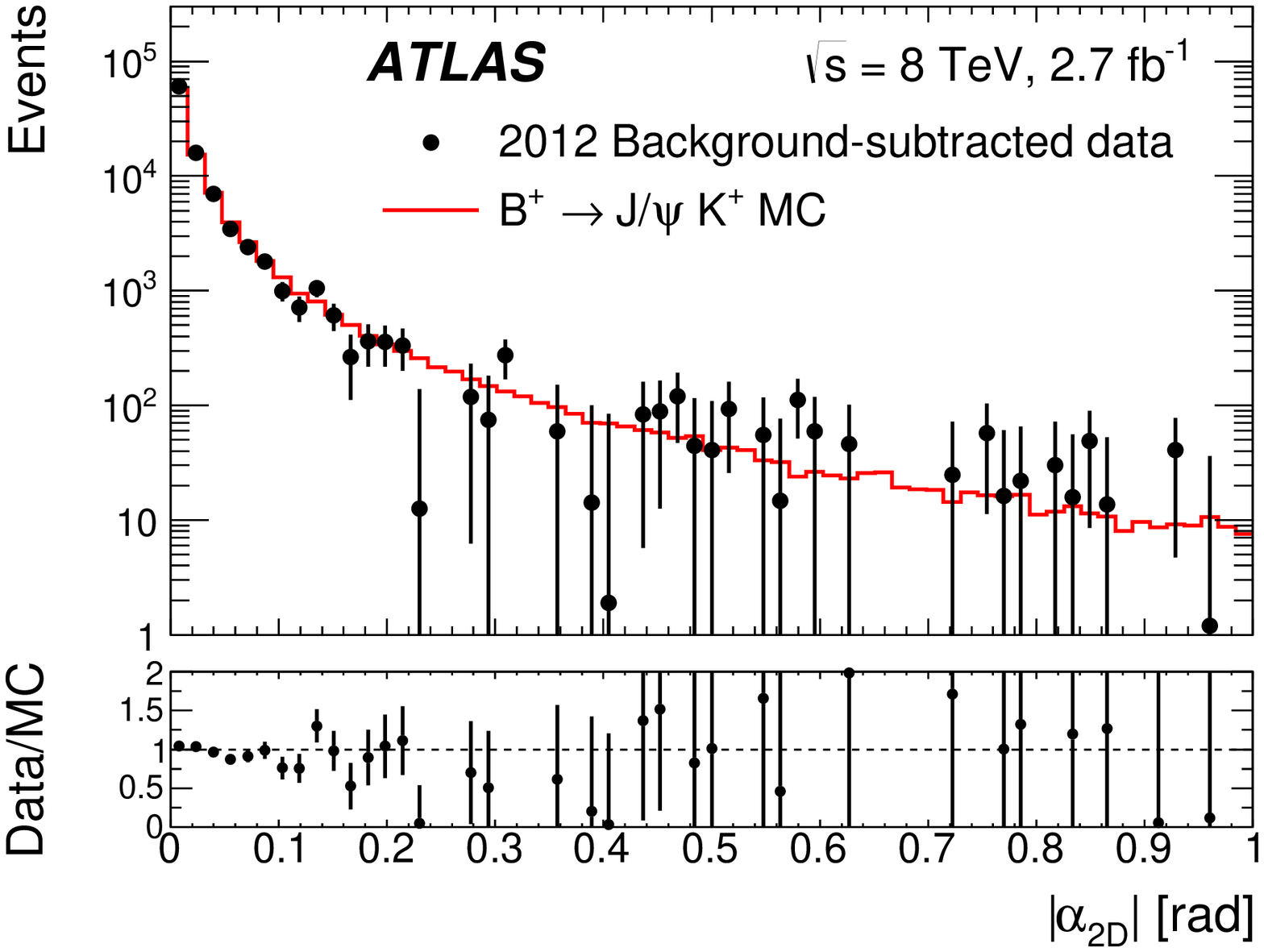}
    }
    \subfigure[]{
      \hspace*{-0.5cm}
      \includegraphics[width=0.52\textwidth]{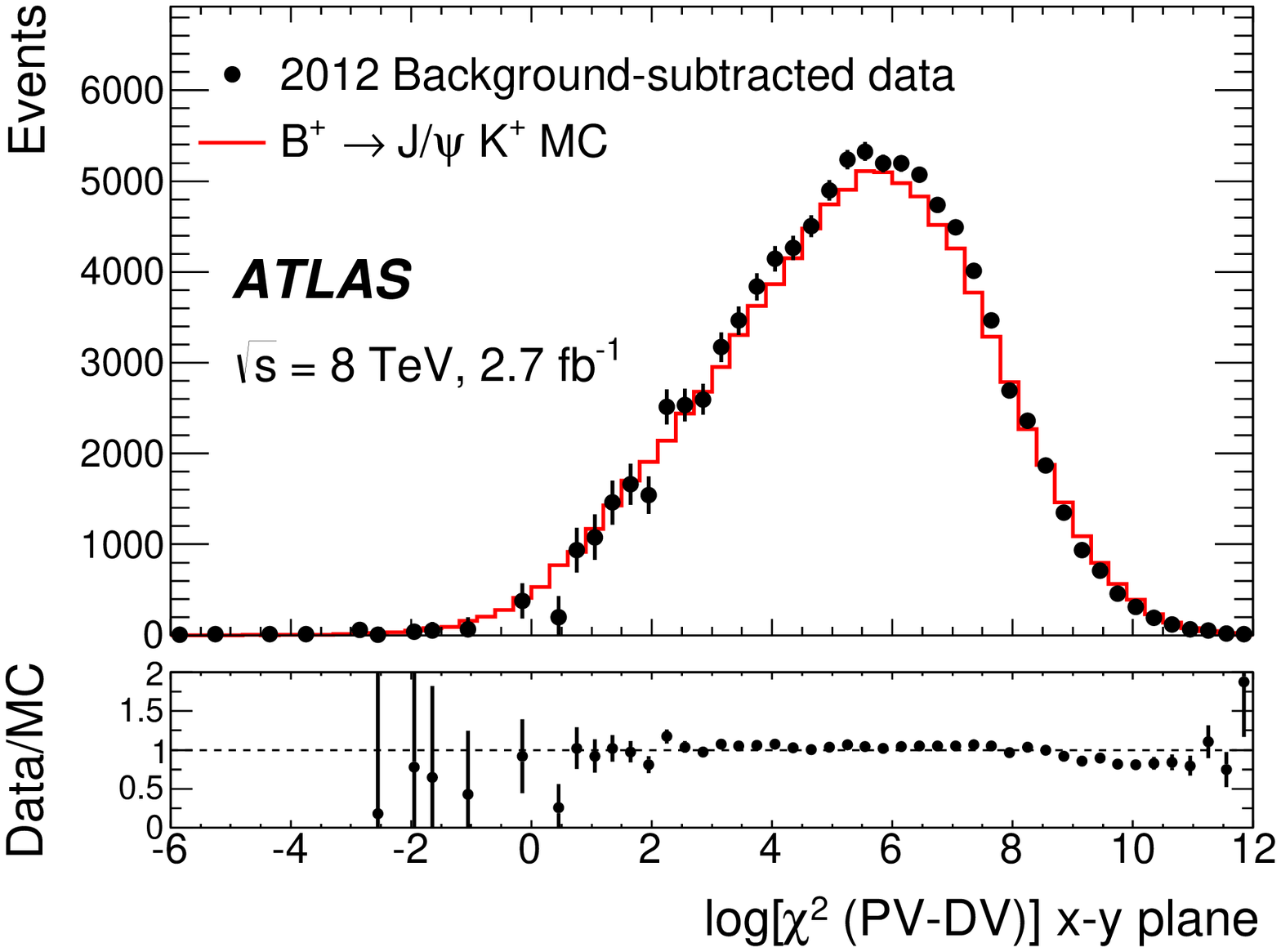}
      \hspace*{-0.6cm}
    }\\
    \subfigure[]{
      \hspace*{-0.5cm}
      \includegraphics[width=0.52\textwidth]{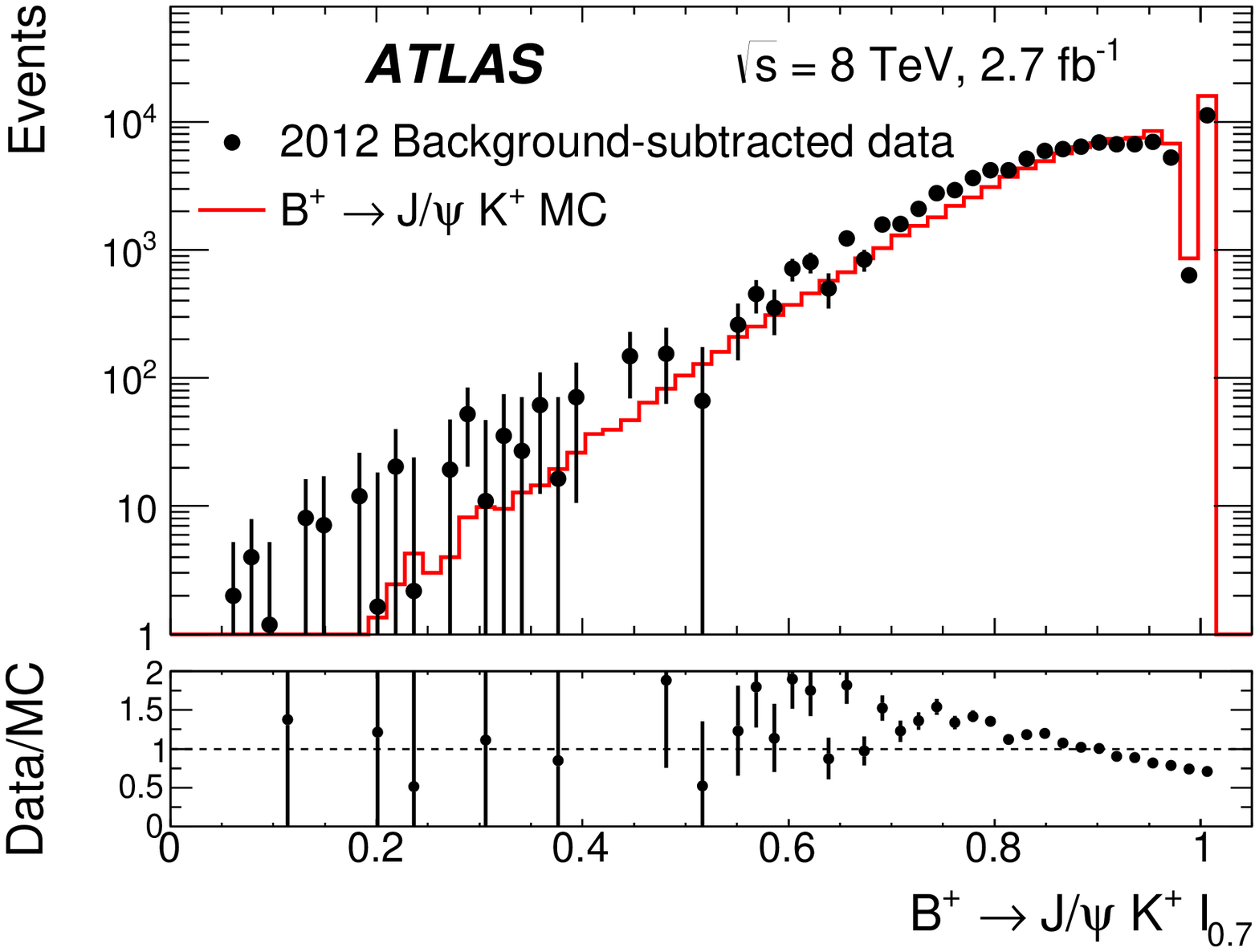}
    }
    \subfigure[]{
      \hspace*{-0.5cm}
      \includegraphics[width=0.52\textwidth]{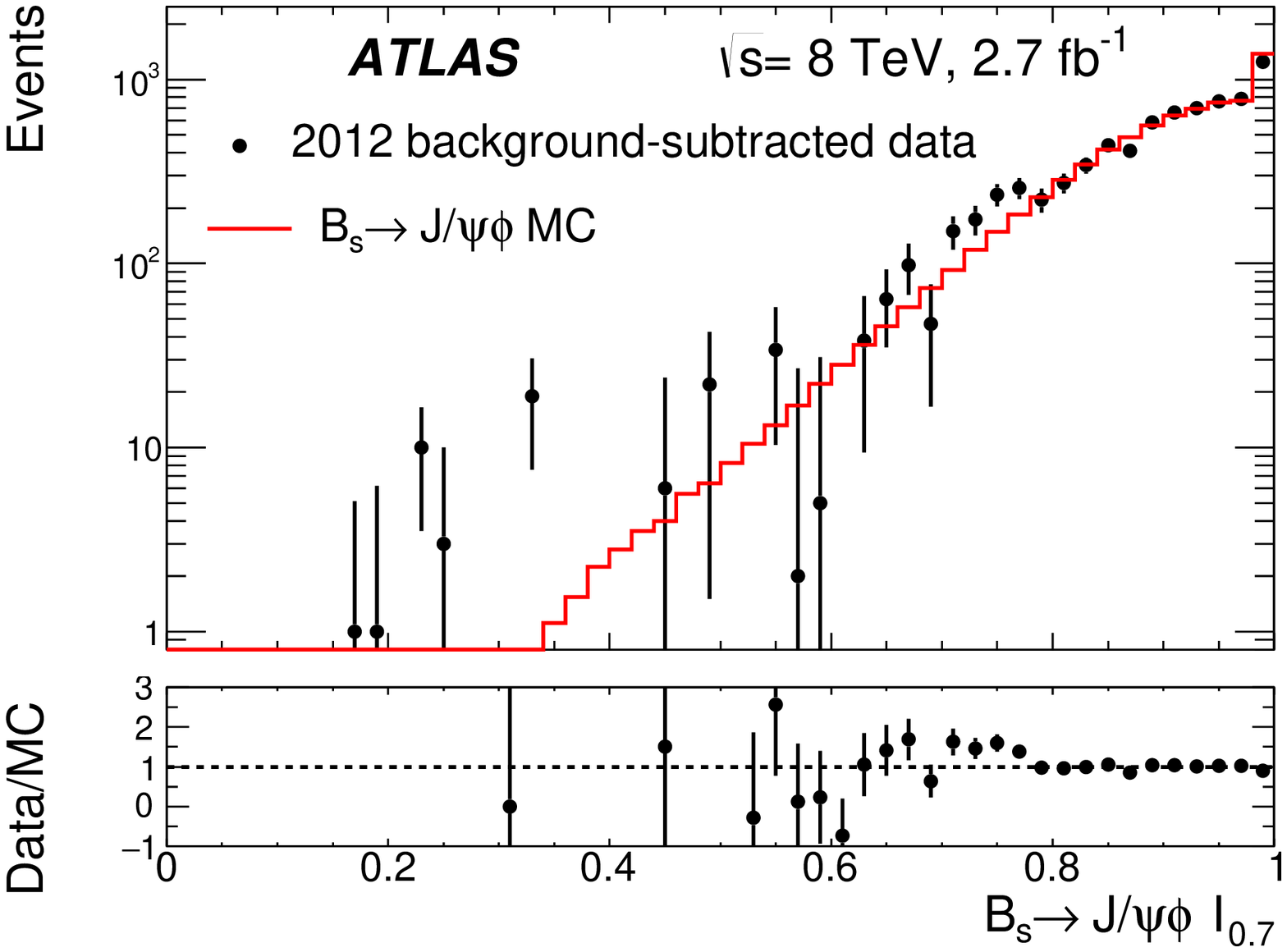}
      \hspace*{-0.6cm}
    }
\caption{
  Data and MC distributions in \BpKpJpsi\ events for 
  the discriminating variables: $|\alpha_{2\mathrm{D}}|$ (a),
  $\chi^{2}{}_{{\mathrm PV,DV}\,xy}$ (b) and $I_{0.7}$ (c).
  The variable $I_{0.7}$ is also shown for \BsJpsiPhi\ events (d).
  The black dots correspond to the sideband-subtracted data, while 
  the red histogram corresponds to 
  the MC distribution, normalised to the number of data events.
  Differences in shape between MC events and data are accounted for as systematic effects. 
  The discrepancy shown for $I_{0.7}$ in the  \BpKpJpsi\ channel
  is the most significant among all variables and both reference channels.
 }
\label{fig:discri-var-comparison}
\end{center}
\end{figure}

The distributions of the discriminating variables are also used for the comparison of 
 \BpKpJpsi\ and \BsJpsiPhi\ events between MC simulation and data. 
To perform such comparison,
for each variable the contribution of the background is subtracted from the signal.
For this purpose, a maximum-likelihood fit is performed to the mass distribution,
separately in the four trigger and data categories.
For \Bp, the signal is described by two overlying Gaussian distributions,
an error function for the partially reconstructed decays
and an exponential function for the continuum background. 
The fit model is simpler than the one used for the extraction of the \Bp\ signal
used for normalisation after the final selection, described in Section~\ref{sec:bplus}, 
but it is sufficient for the purpose discussed here. 
For \BsJpsiPhi, a Gaussian distribution is used for the signal and a third-order Chebychev
polynomial for the background. For each discriminating variable, 
the background distribution observed in the sidebands is interpolated to the 
signal region, normalised according to the result of the likelihood fit, 
and subtracted from the distribution observed in the signal region. 

Figure~\ref{fig:discri-var-comparison} shows examples of the distributions of
the discriminating variables obtained from data and MC simulation.
In general, the overall shapes of distributions are in good agreement between data and
MC events. Observed differences are accounted for as systematic effects 
with the procedure described in Section~\ref{sec:eff}.
The discrepancy shown for the isolation variable $I_{0.7}$ in the  \BpKpJpsi\ channel
is the most significant one among all variables and both reference channels.

%% file: BMuMuPaper_bplus.tex
\section{Yield extraction for the normalisation channel $B^+ \rightarrow J/\psi K^+$}
\label{sec:bplus}

The $B^+$ yield for the normalisation channel is extracted with an
unbinned extended maximum-likelihood fit to the $J/\psi K^+$ mass distribution.
The functional forms used to model both the signal and the backgrounds
are obtained from studies of MC samples. 
All the yields are extracted from the fit to data, while the shape parameters
are determined from a simultaneous fit 
to data and MC samples. Free parameters are introduced for the mass scale and
mass resolution to accommodate data--MC differences. 

\begin{figure}[!htb]
\begin{center}
    \subfigure[]{
      \hspace*{-0.5cm}
      \includegraphics[width=0.52\textwidth]{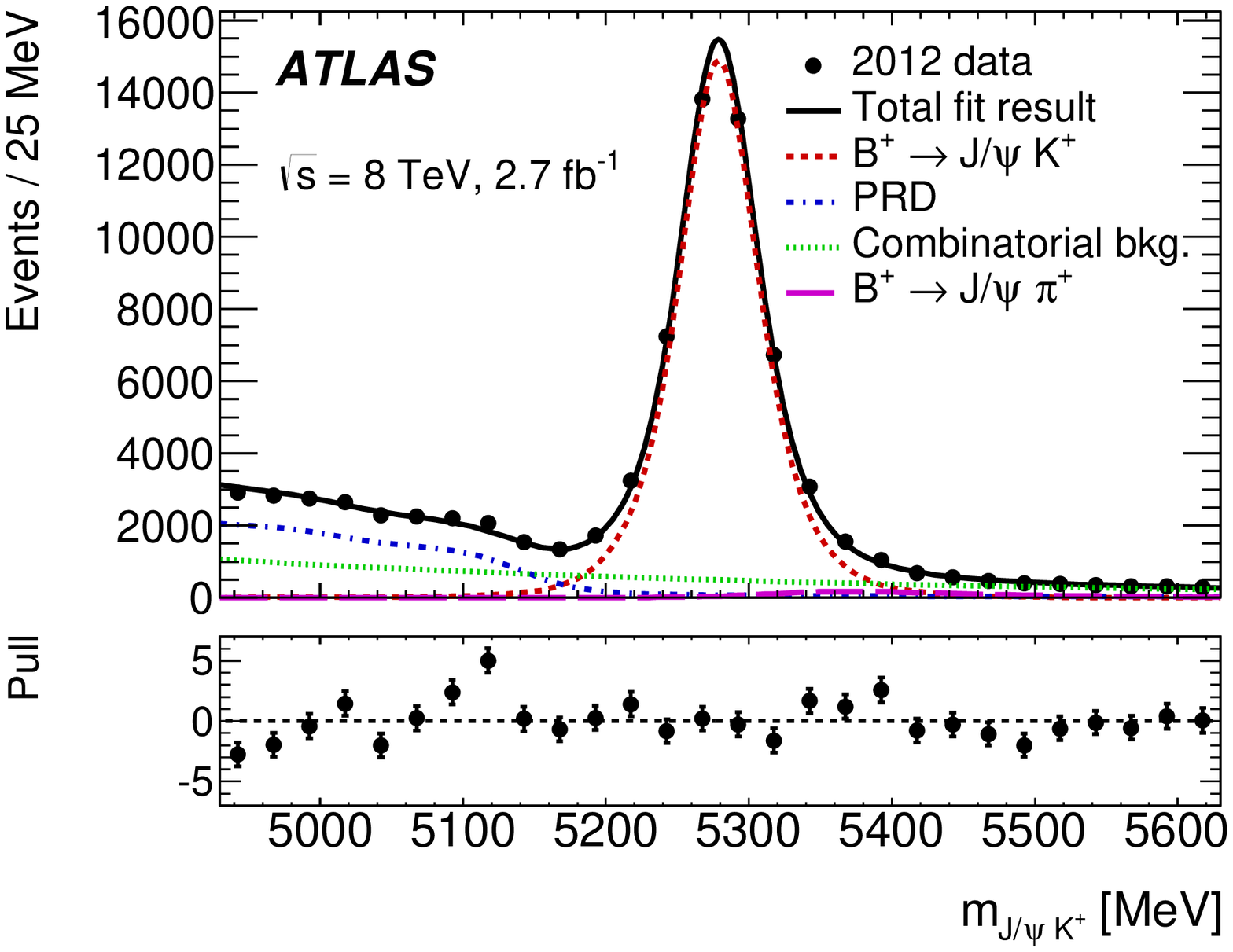}
    }
    \subfigure[]{
      \hspace*{-0.5cm}
      \includegraphics[width=0.52\textwidth]{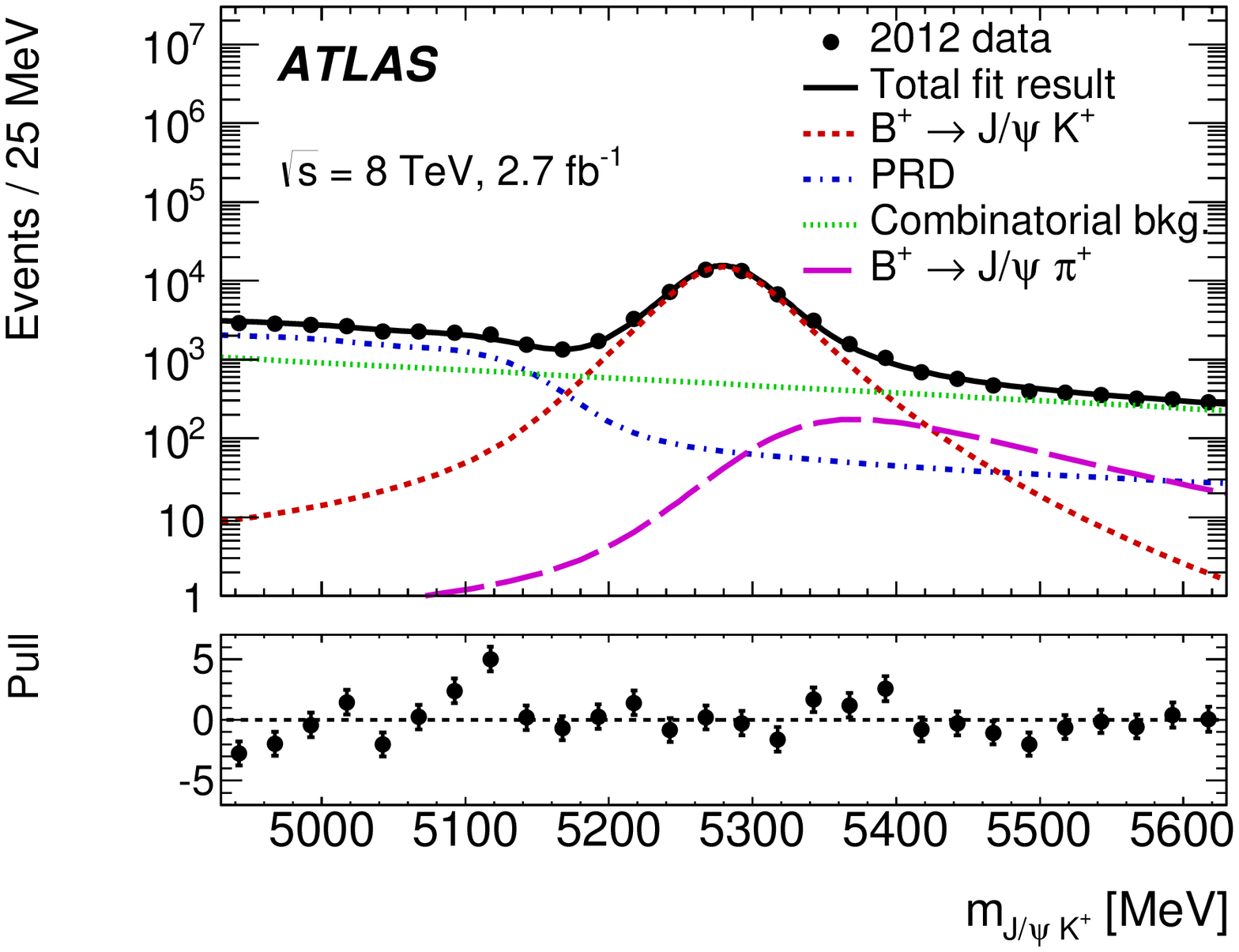}
      \hspace*{-0.6cm}
    }
\caption{$J/\psi K^+$ invariant mass distribution for all $B^+$
candidates in the $T_1$ trigger category in 2012 data in linear (a)
and logarithmic (b) scale. The result of the fit is overlaid.
The various components of the spectrum are described in the text.
The insets at the bottom of the plots show the bin-by-bin pulls for the 
fits, where the pull is defined as the difference between the data point
and the value obtained from the fit function, divided by the error from the fit.
}
\label{fig:bplus}
\end{center}
\end{figure}

The fit includes four components: \BpKpJpsi\ events, Cabibbo-suppressed
\BpPipJpsi\ events on the right tail of the main peak, partially reconstructed
$B$ decays (PRD) where one or more of the final-state particles are missing,
and the combinatorial background composed mostly of $b\bar{b}\to \jpsi X$ events.
The shape of the \BpKpJpsi\ distribution is parameterised using a Johnson $S_U$
function~\cite{Johnson-1,Johnson}
and a  Gaussian function for the $T_1$, $T_2$ and $2011$ categories, while a single
Johnson $S_U$ function is used for the $T_3$ category.
The final \BpKpJpsi\ yield includes the contribution from
radiative decays.
The \BpPipJpsi\ events are modelled by the sum of a Johnson $S_U$ and a Gaussian
function, where all parameters are determined from the MC simulated data.
The PRD are described with combinations of Fermi--Dirac and exponential functions,
slightly different between the different categories in the low-mass region.
Their shape parameters are determined from simulation. 
Finally, the combinatorial background is modelled with an exponential function
with the shape parameter extracted from the fit.
As an example, the fit for the $T_1$ category is shown in Figure~\ref{fig:bplus}.
The results of the fits in all data categories are shown in Table~\ref{tab:bplus}.
\begin{table}[!htb]
  \begin{center}
    \caption{Results of the fits to the events reconstructed as $\BpKpJpsi$ 
      in each trigger and data category. Uncertainties are statistical and systematic,
      respectively. 
    }
    \begin{tabular}{ccc}
      \toprule
      Category     &   $N_{J/\psi K^+}$             &    $N_{J/\psi\pi^+}$          \\
      \midrule
      $T_1$        &   $46\,860 \pm 290 \pm \hphantom{0}\,280$  &    $1\,420 \pm 230 \pm \hphantom{0}\,440$   \\
      $T_2$        &   $\hphantom{0}5\,200 \pm  \hphantom{0}84 \pm \hphantom{0}\,100$  &    $ \hphantom{0}\,180 \pm  \hphantom{0}51 \pm \hphantom{00}\,89$   \\
      $T_3$        &   $\hphantom{0}2\,512 \pm  \hphantom{0}91 \pm \hphantom{00}\,42$  &    $ \hphantom{00}\,85 \pm  \hphantom{0}77 \pm \hphantom{00}\,30$   \\
      $2011$       &   $95\,900 \pm 420 \pm 1\,100$  &    $3\,000 \pm 340 \pm 1\,140$   \\
      \bottomrule
    \end{tabular} 
    \label{tab:bplus}.
  \end{center}
\end{table}

Some of the systematic effects are included automatically in the fit:
the effect of limited MC sample size, for example, is included in the
uncertainties through a simultaneous fit to data and MC samples. 
Scaling factors determined in the fit to data account for the differences in mass scale
and resolution between data and MC simulation.
Additional systematic uncertainties are evaluated by varying the default
fit model described above: they take into account the kinematic differences
between data and the MC samples used in the fit,
differences in efficiency between $B^+$ and $B^-$ decays, 
uncertainties in the relative fractions and shapes of PRD, and in the shape
of the combinatorial background. 
In each case, the difference with respect to the default fit is recorded,
symmetrised and used as an estimate of the systematic uncertainty.
The main contributions to the systematic uncertainty come from the shape of the combinatorial 
background, the relative fractions of PRD and the signal charge asymmetry.
The total statistical and systematic uncertainty in the $B^+$ normalisation
yield amounts to $0.8\%$.

\subsection{$J/\psi \pi^{+} {\Huge /} J/\psi K^{+}$ ratio measurement}
\label{sec:bpluspikratio}

For further validation of the fit to the \BpKpJpsi\ yield, the fit described 
in Section~\ref{sec:bplus} 
is used to extract the yields for
\BpKpJpsi\ and \BpPipJpsi\ 
decays and obtain the 
ratio $\rho_{\pi/K}$ of the corresponding branching fractions.
The measurement is performed separately in the four categories, and combined 
into an error-weighted mean $ {\rho_{\pi/K}}$. 
Table~\ref{tab:bplus} shows the fitted yields.

Most systematic effects cancel in the measurement of this ratio. Residual
systematic uncertainties in the ratio of the branching fractions come from the 
uncertainties in the $K^{-}/K^{+}$, $\pi^{-}/\pi^{+}$ and $K^{+}/\pi^{+}$ 
relative efficiencies.
For each systematic effect the ratio is re-evaluated, therefore accounting
for correlated effects.
The largest systematic uncertainty in the measured ratio comes from the
combinatorial background model parameterisation ($23\%$), followed 
by the effect of the uncertainties in the PRD fraction estimates ($15\%$).
All other systematic sources have uncertainties at the level of $10\%$ or less.
The final result for the ratio of branching fractions is:
\[
 {\rho_{\pi/K}} = \frac{\BR(B^{+} \to J/\psi \pi^{+} )}{\BR(B^{+} \to J/\psi K^{+}) }
= 0.035 \pm 0.003 \pm 0.012 \:,
\]
where the first error is statistical and the second is systematic.
The result is in agreement with the most accurate available results
from LHCb ($0.0383 \pm 0.0011 \pm 0.0007$~\cite{Aaij:2012jw})
and BABAR ($0.0537 \pm 0.0045 \pm 0.0011$~\cite{Aubert:2004pra}).

%% file: BMuMuPaper_eff.tex
\section{Evaluation of the $B^0{}\!_{(s)}\to \mumu$ to \BpKpJpsi\ efficiency ratio}
\label{sec:eff}
The ratio of efficiencies for \BpKpJpsi\ and \Bmumu\ enters the
$\mathcal{D}_{\mathrm{norm}}$ term defined in Eq.~(\ref{eq:DnormFormula}).
Both channels are measured in the fiducial volume of the $B$ meson
defined as $\pt^B>8.0$~\GeV\ and $\left| \eta_B\right|<2.5$.

The efficiencies within the fiducial volume include both acceptance and selection efficiency.
The acceptance is defined by the selection placed on the particles in the final state: 
$\pt^{\mu}>4.0$~\GeV\ and $|\eta_{\mu}|<2.5$ for muons, 
$\pt^{K}>1.0$~\GeV\ and $\left| \eta_{K}\right|<2.5$ for kaons. 
In addition to the reweighting of the distributions of $\pt^B$, $|\eta^B|$ and
the number of reconstructed PVs observed in data, the MC samples are reweighted
according to the equivalent integrated luminosity associated with each trigger
category and the Level-2 muon trigger algorithms used in 2012.

The selection efficiency includes trigger, reconstruction, and signal selection. 
The trigger efficiencies are taken from a data-driven study 
based on the comparison of single-muon and dimuon triggers for events containing
muon pairs from the decays of $J/\psi$ and $\Upsilon$ resonances~\cite{tag-n-probe-Upsilon}.
The signal selection requires the output of the continuum-BDT to be larger than 0.24.

All efficiency terms are computed separately for the three trigger selections
used in 2012 and for the 2011 sample.
Table~\ref{tab:effratio} provides the values of the efficiency ratios
\theRhok, for each of the categories ($k=1-4$), together with the statistical
and systematic uncertainties described below. 

The efficiency ratios shown in Table~\ref{tab:effratio} are computed
using the mean lifetime of $\Bs$~\cite{HFAG2014, pdg14} 
in the MC generator. The same efficiency ratios apply to the \Bsmumu\ and \Bdmumu\
decays, within the MC statistical uncertainty of $\pm 0.5\%$. 

Table~\ref{tab:systematics-2} provides a summary of the systematic uncertainties in 
$\mathcal{D}_{\mathrm{norm}}$.  Each contribution is discussed below.

\begin{table}[!b]
\begin{center}
\caption{Values of the efficiency ratios \theRhok\ 
for the 2012 trigger categories and the 2011 sample,
and their relative contributions to $\mathcal{D}_{\mathrm{norm}}$ 
(Eq.~(\ref{eq:DnormFormula})). 
The first uncertainty is statistical and the second systematic. 
The systematic component includes the uncertainties
from the MC reweighting and from data--MC discrepancies, as described in the text.
The correction due to the $\Bs$ effective lifetime value discussed in the text
is not applied to the numbers shown.
}
\begin{tabular}{ccc}
\toprule
Data category ($k$) & \theRhok\ = \ensuremath{(\varepsilon_{J/\psi K^+}/\varepsilon_{\mu^+\mu^-})_k}
& Relative contribution to $\mathcal{D}_{\mathrm{norm}}$ \\
\midrule
$T_1$              & $0.180 \pm 0.001\,\pm$\,$0.009$ & $68.3\%$ \\ 
$T_2$              & $0.226 \pm 0.004\,\pm$\,$0.014$  & \hphantom{0}$6.0\%$    \\ 
$T_3$              & $0.189 \pm 0.005\,\pm$\,$0.022$ & \hphantom{0}$3.5\%$  \\ 
$2011$             & $0.156 \pm 0.002\,\pm$\,$0.009$  & $22.2\%$   \\ 
\bottomrule
\end{tabular}
\label{tab:effratio}
\end{center}
\end{table} 

The statistical uncertainties in the efficiency ratios come from the finite number of
events available for the simulated samples. 
The systematic uncertainty affecting \theRhok\ comes from four sources.
A first contribution is due to the uncertainties in the DDW and in the 
trigger efficiencies. This term is assessed from pseudo-MC studies, 
performed by varying the corrections within their statistical
uncertainties. The RMS value of the distribution of \theRhok\ obtained from pseudo-MC samples
is taken as the systematic uncertainty.
The uncertainties 
range from $\pm$1\% to $\pm$5\% depending on the category considered, with a combined effect
of $\pm$1.5\% for the data collected in 2012, and $\pm$2.2\% for 2011. 

A second source of systematic uncertainty arises from the differences
between data and MC simulation observed in the distributions of the discriminating variables 
used in the continuum-BDT classifier (Table~\ref{tab:15vars}). 
For each of the $15$ variables, the MC samples for \Bsmumu\ and \BpKpJpsi\  
are reweighted according to the distribution 
observed in  $B^+$  events from the data sample, after background subtraction.
The difference between the efficiency ratio value obtained after each
reweighting and the value without reweighting is taken as the systematic uncertainty
due to mis-modelling of that variable. 
A comparison of data and MC simulation based on the \BpKpJpsi\ channel is preferred to the one with
 \BsJpsiPhi, since the latter is generally found to be statistically consistent but less precise. 
The isolation variable $I_{0.7}$ is computed on charged-particle tracks only, and differences 
between \Bp\ and \Bs\ are expected and were observed in previous studies~\cite{bsmumuv1}. 
For this variable, the reweighting procedure is based on $B^+$ data for the corresponding 
MC sample, and on \BsJpsiPhi\ data for the \Bsmumu\ sample.
The main discrepancy for the 2012 MC samples is observed in the isolation 
variable $I_{0.7}$, both in the normalisation channel 
($-8.9$\% in $\varepsilon_{J/\psi K^+}$ with the weights extracted from the \BpKpJpsi\ channel) 
and in the signal ($-3.8$\% in $\varepsilon_{\mu^+ \mu^-}$ reweighting with \BsJpsiPhi),  
with an effect of $-5.3$\% on the efficiency ratio.
The absolute values of the shifts in the ratio observed for other variables are smaller than 2\%,
with absolute shifts in the efficiencies of either channel smaller than 5\%.
For the 2011 sample 
a similar pattern is observed, but with smaller amplitudes:  
3.2\% shift for $I_{0.7}$ and less than 1\% in any other variable.
Because of the relatively large mis-modelling of $I_{0.7}$, the 2012 MC samples
obtained after reweighting are taken as a reference, thus correcting the
central value of the efficiency ratio.  A systematic error of $\pm 3.2$\%, 
derived from the uncertainty in the reweighting procedure, 
is taken as the
contribution of $I_{0.7}$ to the uncertainty in \theRhok.
The correlation of the variable $I_{0.7}$ with the other discriminating
variables is found to be negligible in the MC samples and in \Bp\ data. 
For all other variables, the shifts observed with the reweighting
procedure are taken as contributions to the systematic uncertainty.
In each category, the total uncertainty is the sum in quadrature of the uncertainties
over all discriminating variables, and is the dominant contribution to the
systematic uncertainties shown in Table~\ref{tab:effratio}.
\begin{table}[!tb]
\begin{center}
\caption{Summary of the systematic uncertainties 
in the $\mathcal{D}_{\mathrm{norm}}$ term of Eq.~(\ref{eq:DnormFormula}).
}
\begin{tabular}{lc}
\toprule
Statistical uncertainty in simulation                & $0.5\%$ \\
$\pt$, $\eta$ reweighting and trigger efficiency     & $1.3\%$ \\
Data to MC discrepancy in discriminating variables   & $4.2\%$ \\
$K^+$ and $\Bp$ reconstruction                       & $3.6\%$ \\
Residual trigger efficiency systematic uncertainty   & $1.5\%$ \\
\midrule
$B^+$ yield                            & $0.8\%$ \\
\midrule
Total uncertainty  & $5.9\%$ \\
\bottomrule
\end{tabular}
\label{tab:systematics-2}
\end{center}
\end{table}

A third source of systematic uncertainty arises
from differences between the \Bsmumu\ and the \BpKpJpsi\ channels
related to the reconstruction efficiency of the kaon track and of the \Bp\ decay vertex.
The corresponding systematic uncertainty is $\pm 3.6$\%.

Finally, a fourth type of systematic uncertainty is related to residual uncertainties in the
trigger efficiencies, which are evaluated through data-driven studies
performed using $J/\psi K^+$ and $\mu^+\mu^-$ candidates. 
The uncertainties cancel to a large extent in the comparison of the signal and
normalisation channels, and the residual systematic uncertainty, averaged across
the categories, is $\pm$1.5\%.

The efficiency ratios enter in Eq.~(\ref{eq:BRFormula}) 
with the $\mathcal{D}_{\mathrm{norm}}$ term defined in Eq.~(\ref{eq:DnormFormula}). 
For each category $k$, the efficiency ratio is multiplied by the number of observed $B^+$
candidates and the trigger prescaling factor. 
The relative contributions of the $T_1$, $T_2$, $T_3$ and $2011$ categories
are shown in Table~\ref{tab:effratio}.
The uncertainties in \theRhok\  are weighted accordingly and combined.
For the trigger categories of the 2012 data sample, the correlations among the
uncertainties due to DDW, trigger efficiency and mis-modelling of
the discriminating variables are taken into account. 
No significant correlation is found between the 2011 and 2012 samples in 
the mis-modelling of the discriminating variables.
The total uncertainty is equal to $\pm 5.9$\%, including 
the relatively small uncertainty in the yield of \BpKpJpsi.
For comparison, the combined effect on $\mathcal{D}_{\mathrm{norm}}$ of all corrections
applied to the simulated data is equal to $+3.2\%$, with the contributions from DDW,
data-driven trigger efficiencies and $I_{0.7}$ reweighting equal
to $+3.6\%$, $+3.9\%$ and $-4.1\%$, respectively.

A correction to the efficiency ratio for \Bsmumu  is expected because of the width difference 
$\Delta \Gamma_s$ between the \Bs\ eigenstates.
According to the SM, the decay \Bsmumu\ proceeds predominantly through the heavy state 
$B_{s,\mathrm{H}}$~\cite{DeBruyn:2012wk}, which has width
$\Gamma_{s,\mathrm{H}}= \Gamma_s - \Delta \Gamma_s/2$, i.e.~$(6.2\pm0.5)$\% smaller than the 
average $\Gamma_s$~\cite{HFAG2014}.
The variation in the value of the \Bsmumu\ mean lifetime was tested with MC simulation, 
and found to change the \Bs\ efficiency by $+4$\%, and consequently the
\Bs\ to \Bp\ efficiency ratio.
This correction is applied to the central value of $\mathcal{D}_{\mathrm{norm}}$ used
in Section~\ref{sec:br} for the determination of \BR(\Bsmumu). 
Due to the small value of $\Delta \Gamma_d$, no correction needs to be applied to
the \Bdmumu\ decay.

\subsection{Comparison of normalisation yields with other measurements}
\label{sec:sub:yield-check} 
The systematic errors in acceptance and efficiency are minimised by using 
 \BpKpJpsi\ as the normalisation channel
and evaluating  only ratios of efficiencies. 
However, event counts and absolute efficiency values for the reference channels can be used
to extract the production cross sections for the purposes of comparisons with other
measurements. The relevance of such comparisons is mainly in the validation of the
selection efficiency in the analysis presented here, in particular for the final
selection performed with the continuum-BDT classifier. 

The yield of $B^+$ can be compared to the measurement performed by ATLAS
with $2.7$~fb$^{-1}$ of data collected at $\sqrt{s}= 7$~\TeV~\cite{ATLAS:2013cia}, 
and based on the same decay channel. In the comparison, the data collected at
$\sqrt{s}= 8$~\TeV\ for the present analysis were restricted
to the phase space $\pt^B>9.0$~\GeV\ and $\left| \eta_B\right|<2.25$ used for
the previous result. Trigger and preliminary selections are very similar,
but the selections against
continuum background and fake muons are used only in the present analysis. 
The difference in the collision energy is taken into account by comparing
the measured production cross section to the prediction based on the
fixed-order next-to-leading-log (FONLL) approximation~\cite{Cacciari:2012ny}.
The theoretical uncertainty in the extrapolation from $7$ to $8$~\TeV\ is expected
to be small compared to experimental uncertainties. 
The ratio 
of the observed to the predicted cross section was measured in Ref.~\cite{ATLAS:2013cia} 
as $1.24\pm0.04\pm0.09$,
where only measurement uncertainties, respectively statistical and systematic, are
included in the error. The corresponding value from the present analysis
is $1.17\pm 0.02\pm0.14$,
with the uncertainty dominated by the systematic error in the efficiency of the continuum-BDT selection. 
The result is in agreement with the previous measurement. 
Correlated systematic errors between the two analyses amount to $\pm 0.05$. 

The measurements of \BsJpsiPhi\ and \BpKpJpsi, together with the corresponding
acceptance and efficiency values, 
can be used to extract the production ratio $\Bs/\Bp$,
for $10 \lesssim \pt^B \lesssim 20$~\GeV\ and $|\eta|<2.5$, in $pp$
collisions at $\sqrt{s}= 8$~\TeV. 
Using world averages values~\cite{pdg14} for the branching fractions to the
final states, the resulting mean ratio of the hadronisation fractions $f_s/f_u$
is equal to $0.236 \pm 0.014 \pm 0.018 \pm 0.021$,
where the first error is statistical, the second is the systematic uncertainty
in the efficiency ratio 
and the third is the uncertainty in the branching fractions. 
The ratio is uniform across the kinematic range observed,
and it varies by only $-2\%$ if the $\Bs$ and $\Bp$ signals are extracted
without applying the continuum-BDT selection. 
The normalisation procedure might not be free of bias, 
since the value of $\BR(\BsJpsiPhi)$ includes assumptions about $f_s$, 
and updating the assumptions may change it by about $5\%$. 
The result nevertheless provides a satisfactory consistency check 
with the available measurements~\cite{Aaij:2013qqa,Aad:2015cda}.
The most direct comparison is with the recent value  $f_s/f_d = 0.240 \pm 0.020$~\cite{Aad:2015cda}, 
obtained by ATLAS from the analysis of $2.7$~fb$^{-1}$ of data collected at $\sqrt{s}= 7$~\TeV,  
and performed over the same $\pt^B$ and $\eta^B$ ranges used in this analysis. The 
uncertainty in that measurement is dominated by the prediction of the ratio of 
branching fractions $\BR(\BsJpsiPhi)\,/\,\BR(\Bz \to J/\psi K^{*0})$. 
The ratio of the efficiency-corrected event yields observed at $\sqrt{s}= 8$~\TeV\ in the
present analysis can be compared to the corresponding value from Ref.~\cite{Aad:2015cda}
after rescaling by the ratio of branching fractions 
$\BR(\BpKpJpsi)\, /\, [\BR(B^0 \to J/\psi K^{*0})\!\times\! \BR(K^{*0}\to K^+ \pi^-)]$,
which is known with better accuracy than 
$\BR(\BsJpsiPhi)$. 
In this way, some systematic uncertainties are removed,
and the ratio of the two results is $0.96\pm0.12$.
The largest contribution to the systematic uncertainty is from the efficiency 
of the continuum-BDT selection used in the present analysis. 

In conclusion, the observed event rates for the normalisation channels
are in agreement with previous measurements within uncertainties
of about $12\%$.

%% file: BMuMuPaper_sigfit.tex
\section{Extraction of the signal yield}
\label{sec:sigfit}
Dimuon candidates passing the preliminary selection and the multivariate selections
against hadron misidentification and continuum background are classified according to 
three intervals in the continuum-BDT output: $0.240$--$0.346$, $0.346$--$0.446$ and $0.446$--$1$.
Each interval corresponds to an equal efficiency of $18\%$ for signal events,
and they are ordered according to increasing signal-to-background ratio.
In each continuum-BDT interval, events from the four trigger and data categories are merged.

An unbinned extended maximum-likelihood fit is performed on the dimuon invariant mass distribution
simultaneously across the three continuum-BDT intervals.
The result of the fit is the total yield of \Bsmumu\ and \Bdmumu\ events in the three BDT intervals.
The parameters describing the background are allowed to vary freely and are determined by the fit.
The fit model for signal and background is described in Section~\ref{sub:sec:model}.
The systematic uncertainties related to the BDT intervals, 
to the signal and to the background model are discussed in 
Sections~\ref{sub:sec:model} and \ref{sub:sec:fitsyst}, and
are included in the likelihood with Gaussian multiplicative factors
with width equal to the systematic uncertainty. 

\subsection{Signal and background model}
\label{sub:sec:model}

The model for describing signal and background is based on simulation and on data
collected in the mass sidebands of the search region. 

\begin{figure}[!tbh]
  \begin{center}
  \includegraphics[width=0.52\textwidth]{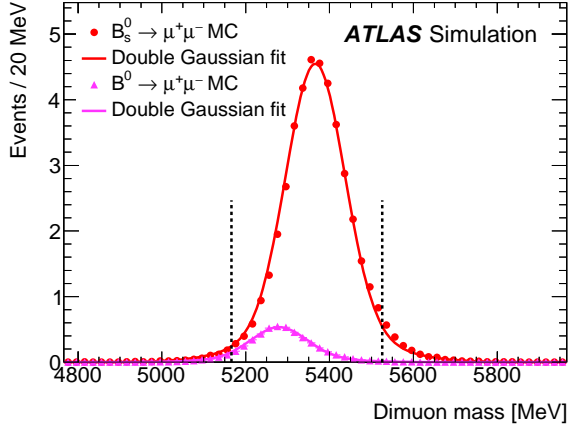}
  \caption{Dimuon invariant mass distribution for the \Bs\ and \Bz\ signals
    from MC simulation. The double Gaussian fits are overlaid.
    The two distributions are normalised to the SM prediction for the expected yield with
    an integrated luminosity of 25 fb$^{-1}$.  
    }
  \label{fig:masssig}
  \end{center}
\end{figure}

The mass distribution of the \Bmumu\ signal is described by a superposition of
two Gaussian distributions, both centred at the \Bz\ or \Bs\ mass.
The parameters are extracted from MC simulation, and they are taken to be uncorrelated with the BDT output.
Systematic uncertainties in the mass scale and resolutions are considered separately.
Figure~\ref{fig:masssig}
shows the mass distributions for \Bz\ and \Bs, obtained from MC events and normalised to
the SM expectations.

The efficiency of the three intervals in the continuum-BDT output for \Bmumu events 
is calibrated with studies performed on the reference channels.
The distribution of the BDT output is compared between MC and background-subtracted data. 
The differences observed in the ratio of data over simulation are described with a linear 
dependence on the BDT output.  
The slopes are equal within $\pm 12$\% between \BpKpJpsi\ and \BsJpsiPhi\,  
and the mean value is used to reweight the BDT-output distribution in the \Bmumu\ MC sample. 
The corresponding variation in the efficiencies are 
equal to $+0.018$ and $-0.018$ respectively in the first and third BDT intervals. 
The values of the lower edge of the second and third BDT intervals are corrected 
in MC simulation to obtain equal efficiencies of $0.180$ in each interval.

The systematic uncertainties in the efficiency of the BDT intervals are obtained 
with a procedure similar to the one used for the event selection (Section~\ref{sec:eff}).
For each discriminating variable, the MC sample is reweighted according to the 
difference between simulation and data observed in the reference channels. 
The variation in the efficiency of each BDT
interval is taken as the contribution to the to the systematic uncertainty due to
mis-modelling of that variable. 
In each BDT interval, the sum in quadrature
of the variations of all discriminating variables is found to be similar in the \BpKpJpsi\ and \BsJpsiPhi\ channels, 
and the average of the two is taken as the total systematic uncertainty in the efficiency.
Values of $\pm0.026$, $\pm0.010$ and $\pm 0.023$ are found respectively in the first, second and third interval.  
Gaussian terms are included in the likelihood in order to describe these uncertainties,
taking care of constraining the sum of the efficiencies of the three intervals,
since that  uncertainty is already included in the selection efficiency.

\begin{figure}[!bth]
  \begin{center}
    \subfigure[]{
      \hspace*{-0.5cm}
      \includegraphics[width=0.52\textwidth]{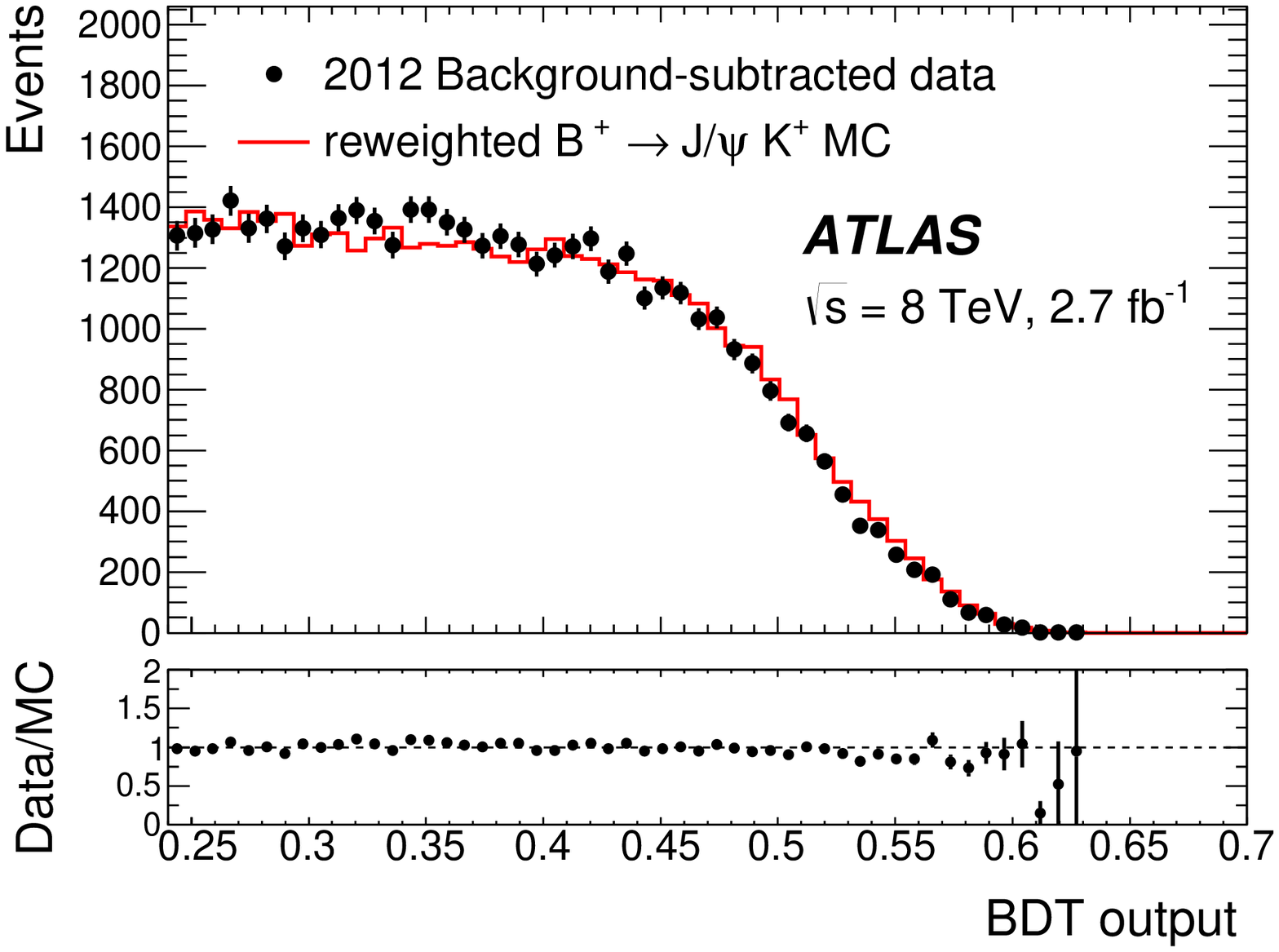}
    }
    \subfigure[]{
      \hspace*{-0.5cm}
      \includegraphics[width=0.52\textwidth]{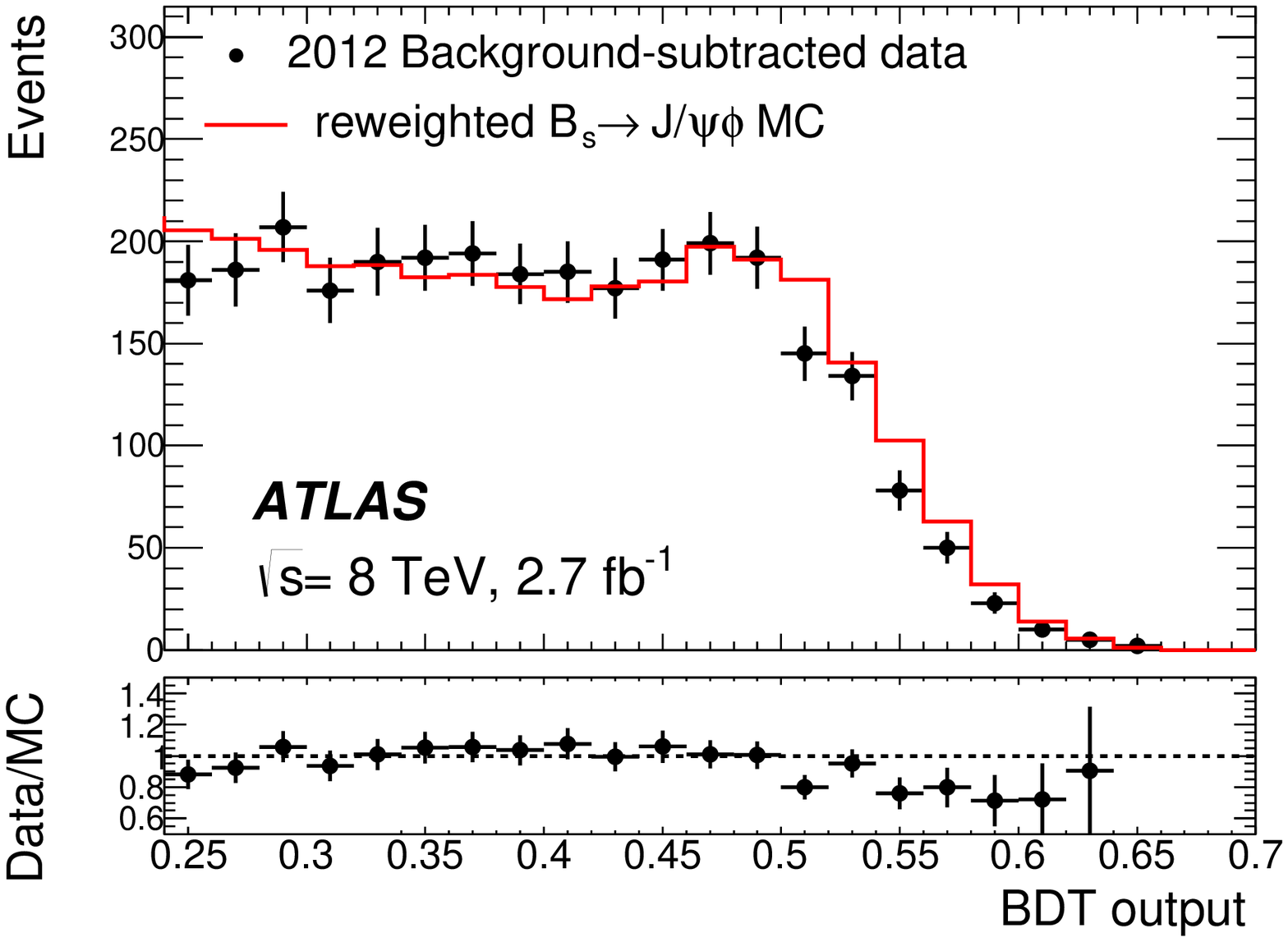}
      \hspace*{-0.6cm}
    }\\
    \subfigure[]{
      \hspace*{-0.5cm}
      \includegraphics[width=0.52\textwidth]{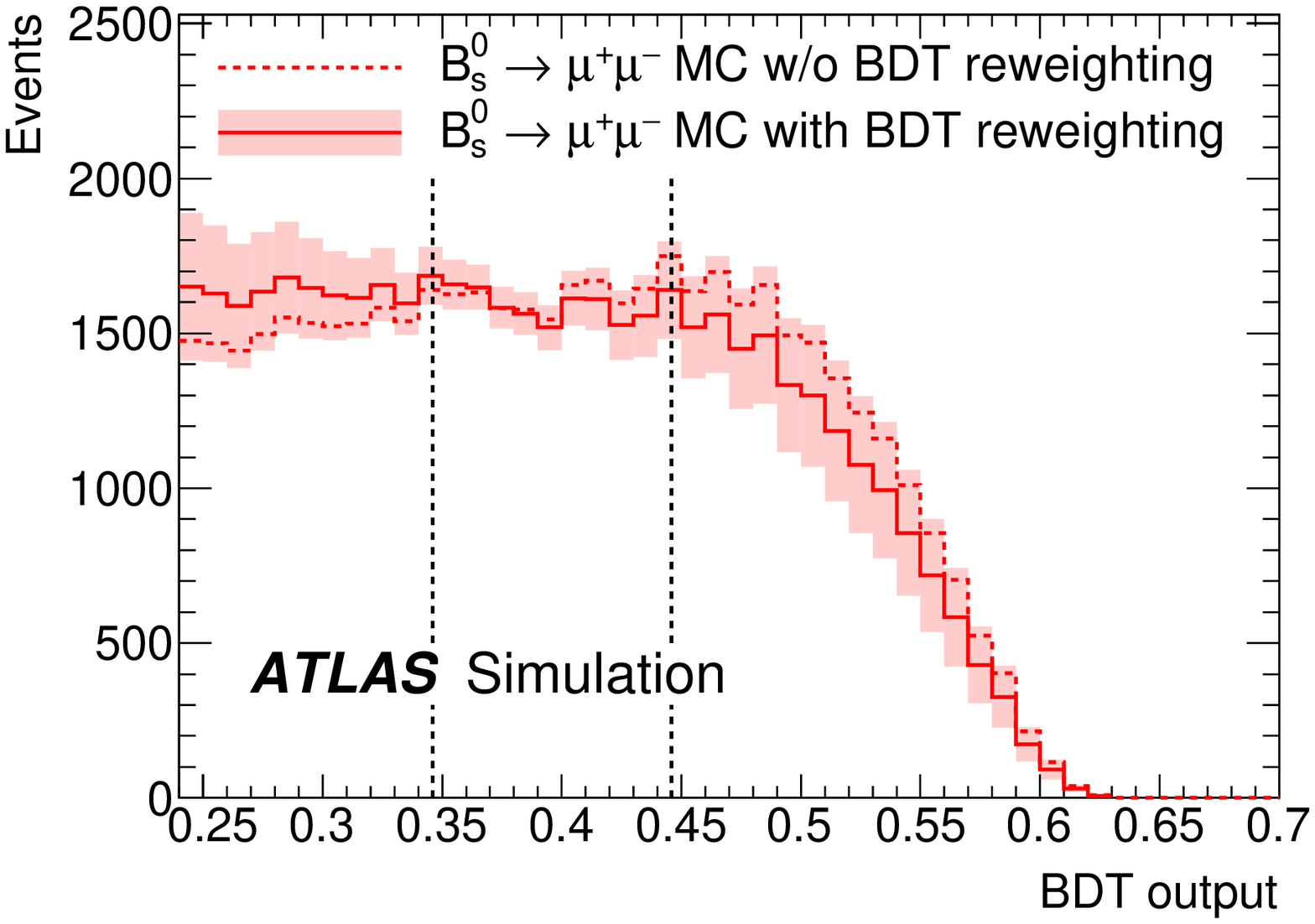}
    }
    \caption{Data and MC distributions of \BpKpJpsi\ (a) and \BsJpsiPhi\ (b)
      for the continuum-BDT output and the MC distributions for \Bsmumu\ (c). 
      The MC samples are normalised to the number of data events. A linear correction
      has been applied to the MC distributions, equal for all channels, and a systematic uncertainty is assigned
      to the distribution of the \Bmumu\ MC sample, as discussed in the text 
      and illustrated by the dashed line and the envelope shown in (c).
      The vertical dashed lines in (c) correspond to the boundaries of the continuum-BDT
      intervals used for the signal extraction.}
   \label{fig:BDTzoomed}
  \end{center}
\end{figure}

Figure~\ref{fig:BDTzoomed} shows the distribution of the continuum-BDT output from data and simulation 
for the reference channels, after reweighting the MC sample.  The MC distribution for
\Bmumu\ events is also shown, illustrating the correction based on the 
BDT output and the systematic uncertainty discussed above. 
The reweighting on the $I_{0.7}$ variables, discussed in Section~\ref{sec:eff} for the 
evaluation of the efficiency of the final event selection (BDT output $> 0.24$), is not applied to the events shown 
in Figure~\ref{fig:BDTzoomed}, and in the evaluation of the relative efficiency of the intervals used for the
extraction of the \Bmumu signal.  Reweighting the BDT output is preferred over reweighting $I_{0.7}$, 
because of correlations present between the discriminating variables after the final selection is applied.

Finally, for the \Bsmumu\ signal, the lifetime difference between $B_{s,\mathrm{H}}$
and \Bs\ requires further corrections to the efficiency of the BDT intervals
of $+0.003$ and $+0.018$ respectively in the second and third interval.
 
The background is composed of the types of events described in Section~\ref{sec:bkg}:
(a) the combinatorial background, which is strongly reduced by the selection progressively
applied in the three intervals of continuum-BDT; 
(b) the background from partially reconstructed SS and SV events,
which is present mainly in the low-mass sideband;
(c) the peaking background.  

The dependence of the combinatorial background on the dimuon mass is described with a first-order polynomial.
In the simulation, the slope of the distribution is similar in the three continuum-BDT intervals.
The correlation between continuum-BDT and dimuon mass is small,
and similar between simulation and sideband data within large statistical uncertainties.
Hence the slope of the mass dependence is described by independent 
parameters in the three intervals, subject to loose Gaussian constraints of uniformity 
within $\pm 40\%$ between the first and second interval, and $\pm 80\%$ between the
first and the third. Such variations of slope are larger than those observed in
simulation, and consistent with those determined from data. 
Deviations from these assumptions are discussed below in Section~\ref{sub:sec:fitsyst}.
The normalisation of the combinatorial background is also extracted independently in each BDT interval.

The SS+SV background has a mass distribution peaking below the low-mass sideband region. 
The mass dependence is derived from data in the low-mass sideband region, and
described with an exponential function with equal shape in the three continuum-BDT intervals.  
The value of the shape parameter is extracted from the fit to data. The normalisation values are
extracted independently in each interval.

The mass distribution of the peaking background is very similar to the \Bz\ signal,
as shown in Figure~\ref{fig:SS-SV}(b). In the fit, this contribution
is included with fixed mass shape and with a normalisation of $1.0 \pm 0.4$ events,
as discussed in Section~\ref{sec:fakes}. 
This contribution is equally distributed among the three intervals of continuum-BDT.

The fitting procedure is tested with pseudo-MC experiments, as discussed below.
The use of three intervals in the continuum-BDT output is found to optimise the performance of the likelihood fit,
with all BDT intervals contributing to the determination of the background, while the second and in particular
the third interval provide sensitivity to the signal yield.

\subsection{Systematic uncertainties in the fit}
\label{sub:sec:fitsyst}
Studies based on pseudo-MC experiments are used to assess the sensitivity
of the fit to the input assumptions.
Variations in the description of signal and background components are used 
in the generation of the pseudo-MC samples.
The corresponding deviations in the average 
numbers $N_s$, $N_d$ of \Bs\ and \Bz\ events returned by the fit,
run in the nominal configuration,  are taken as systematic uncertainties.
The amplitude of the variations in the generation of the pseudo-MC samples 
is determined in some cases by known characteristics of the ATLAS detector
(reconstructed momentum scale and momentum resolution), in others using MC evaluation 
(background due to semileptonic three-body $B^0{}\!_{(s)}$ decays and to $B_c \to J/
\psi \mu$), and in others from uncertainties determined from data in the sidebands
and from simulation (shapes of the background components and their variation across
the continuum-BDT intervals).

The pseudo-MC experiments were generated with the abundance of the background components
extracted from data in the mass sidebands of the signal region. 
The signal was generated with different configurations, corresponding to the SM prediction,
to smaller values of \BR(\Bsmumu) and to smaller/larger values of \BR(\Bdmumu).

For all variations in the assumptions and all configurations of the signal amplitudes,   
the distributions of the differences between results and generated values, divided by the fit errors 
(pull distributions), are found to be correctly described by Gaussian 
functions with widths approximately equal to one and  values of the mean
smaller than $0.2$ for \Bs\ and smaller than $0.4$ for \Bz.
The distributions obtained from 
pseudo-MC samples generated according to the nominal fit model are used to evaluate fit biases.
For \Bs\  the fit bias is negligible. For \Bz\ the absolute value is smaller than 25\% of the fit error, 
and it is included as an additional systematic uncertainty.

The shifts in $N_s$ or $N_d$  are combined by considering separately
the sums in quadrature of the positive and negative shifts
and taking the larger as the symmetric systematic uncertainty. 
For \Bs, the total systematic uncertainty is found to increase with the assumed size 
of the signal, with a dependence
$\sigma_\mathrm {syst} (N_s) = \sqrt{2^2 +  (0.06\times  N_s)^2}$. 
The total systematic uncertainty for \Bz\ is approximately  $\sigma_\mathrm {syst} (N_d) = 3$. 
Most of the shifts observed have opposite sign for $N_s$ and $N_d$, resulting
in a combined correlation coefficient in the systematic errors  
of $\rho_\mathrm {syst}=-0.7$. 

The fit to the yield of \Bs\ and \Bz\ events is modified by including in the likelihood
two smearing parameters for $N_s$ and $N_d$  
that are constrained by a combined Gaussian distribution parameterised by the values of 
$\sigma_\mathrm {syst} (N_s)$, $\sigma_\mathrm {syst} (N_d)$ and  $\rho_\mathrm {syst}$.

\begin{figure}[!htb]
  \begin{center}
    \subfigure[]{
      \hspace*{-0.5cm}
      \includegraphics[width=0.52\textwidth]{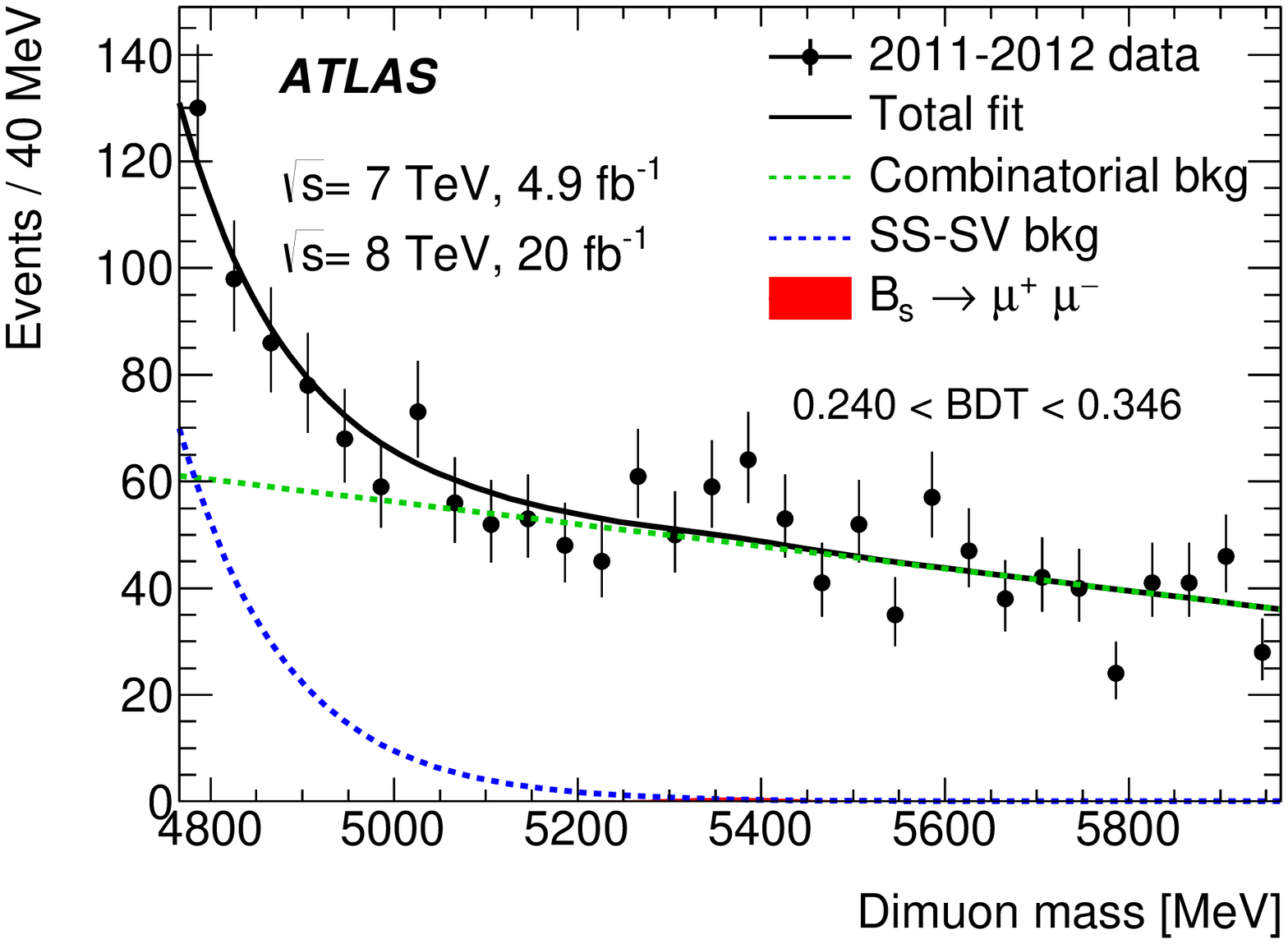}
    }
    \subfigure[]{
      \hspace*{-0.5cm}
      \includegraphics[width=0.52\textwidth]{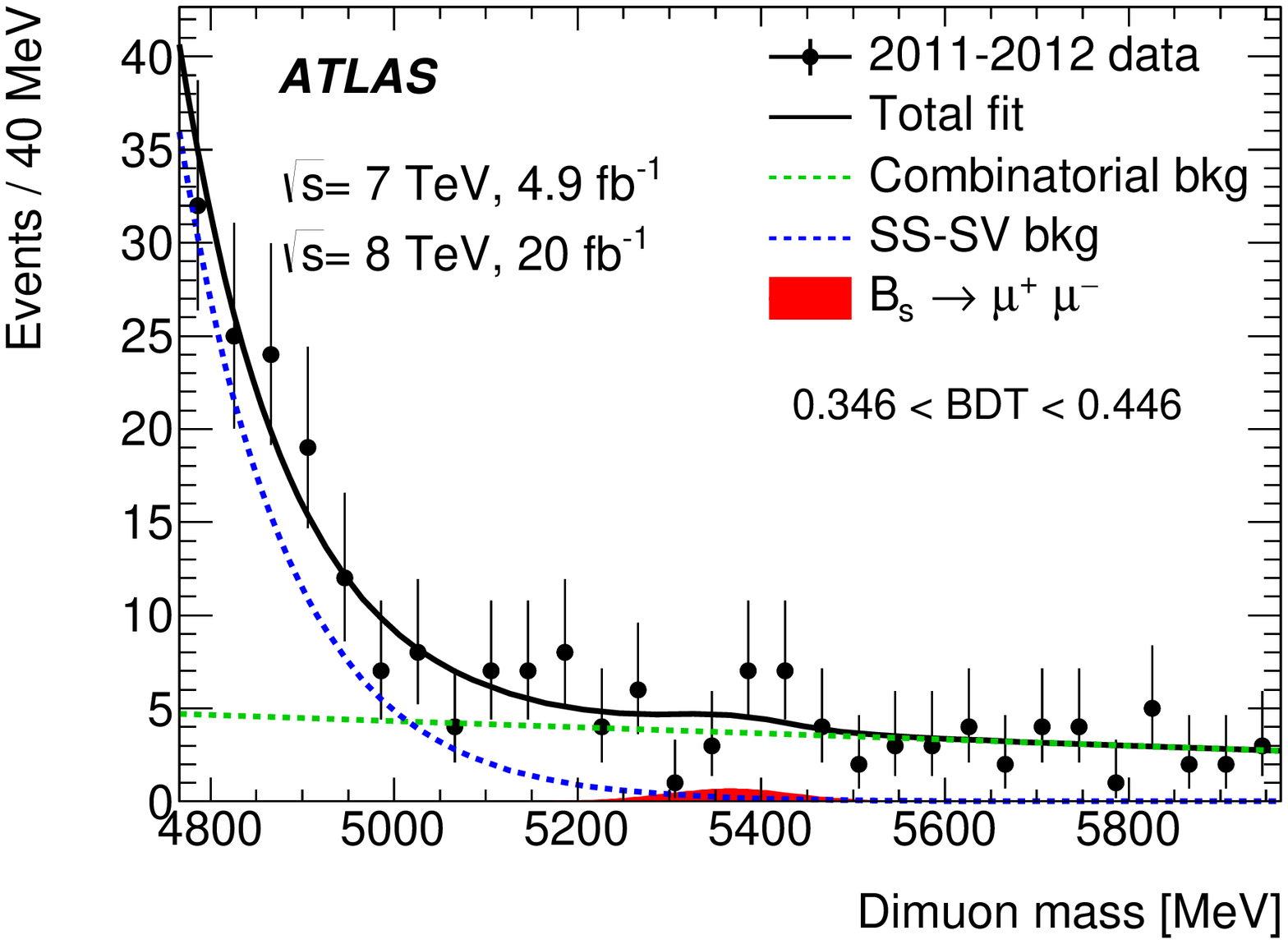}
      \hspace*{-0.6cm}
    }\\
    \subfigure[]{
      \hspace*{-0.5cm}
      \includegraphics[width=0.52\textwidth]{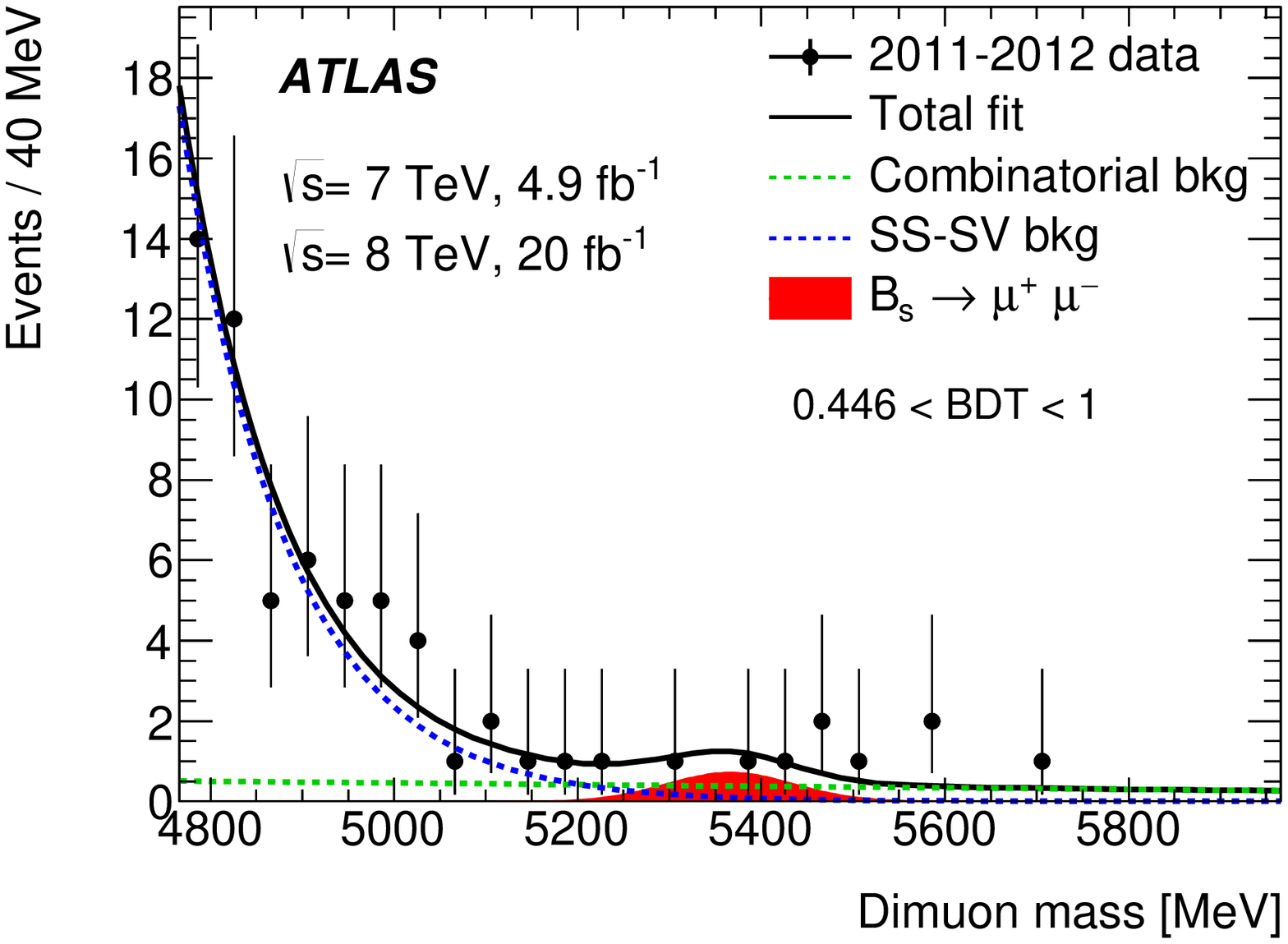}
    }
    \caption{
      Dimuon invariant mass distributions in the unblinded data, 
      in the three intervals of continuum-BDT output. Superimposed is the
      result of the maximum-likelihood fit, obtained imposing the
      boundary of non-negative signal contributions.  The total fit is
      shown as a black continuous line, the filled area corresponds
      to the observed signal component, the blue dashed line to the SS+SV background, 
      and the green dashed line to the combinatorial background.}
  \label{fig:unblinded_default}
  \end{center}
\end{figure}

\subsection{Results of the signal yield extraction}
\label{sec:unblinded}

Including both the $2012$ and $2011$ data-taking periods, 
the numbers of background events contained in the signal region ($5166$--$5526$~\MeV) 
are computed from the interpolation of the data observed in the sidebands. 
The values $509\pm28$, $32\pm6$ and $4.8\pm1.9$ events are obtained respectively
in the three intervals of continuum-BDT. For comparison, the total expected number of
signal events according to the SM prediction is $41$ and $5$ respectively
for $N_s$ and $N_d$,
equally distributed among the three intervals.\footnote{The values of the single-event
sensitivity are discussed in Section~\ref{sec:br}.} 

Once the signal region is made accessible, a total of $1951$ events in the full mass range of 
$4766$--$5966$~\MeV are used for the likelihood fit to signal and background.
Without applying any boundary on the values of the fitted parameters, 
the values determined by the fit are 
$N_s = 16 \pm 12$ and $N_d = - 11 \pm 9$,
where the uncertainties correspond to likelihood variations of $- 2\,\Delta \ln (L) = 1$. 
The likelihood includes the systematic uncertainties discussed above, but
statistical uncertainties largely dominate.  
The primary result of this analysis is obtained by
applying the natural boundary of non-negative yields,
for which the fit returns the values  $N_s = 11$ and $N_d = 0.$
The uncertainties in the result of the fit are discussed in Section~\ref{sec:br},
where the measured values of the branching fractions are presented.

Figure~\ref{fig:unblinded_default} shows the dimuon mass distributions 
in the three intervals of continuum-BDT, together with
the projections of the likelihood fit.

For comparison, the value $N_d$ can be constrained according to the SM
expectation for the ratio $\BR(\Bdmumu)/\BR(\Bsmumu)$ \cite{Bobeth:2013uxa} multiplied
by the ratio of the hadronisation probabilities $f_d/f_s$ \cite{Aad:2015cda},
rather than being extracted independently from the fit.
In this case the value of $N_s$ changes by $-0.8$,
while $N_d = N_\mathrm {s}/8.3 \approx 1.2$.

%% file: BMuMuPaper_br.tex
\section{Branching ratio extraction}
\label{sec:br}

The branching fractions for the decays \Bsmumu\ and \Bdmumu\  are extracted from data 
using a profile-likelihood fit.  The likelihood is obtained from the one
used for $N_s$ and $N_d$ replacing the fit parameters 
with the corresponding branching fractions divided by normalisation terms  
in Eq.~(\ref{eq:BRFormula}), and including Gaussian multiplicative factors 
for the normalisation uncertainties.

The normalisation terms include external inputs for the $\Bp$ branching fraction
and the relative hadronisation probability.
The first is obtained from world averages~\cite{pdg14} as the product of
$\BR(\BpKpJpsi)=(1.027 \pm 0.031) \times 10^{-3}$ 
and $\BR(\Jpsi\to\mumu)=(5.961 \pm 0.033)\%$. 
The second is equal to one for $\Bz$, while for $\Bs$ it is taken from the ATLAS measurement 
$f_s/f_d = 0.240 \pm 0.020$~\cite{Aad:2015cda}, assuming $f_u/f_d=1$~\cite{HFAG2014}.

The efficiency- and luminosity-weighted number of events for the normalisation
channel enters in Eq.~(\ref{eq:BRFormula}) with the denominator $\mathcal{D}_{\mathrm{norm}}$
(Eq.~(\ref{eq:DnormFormula})). 
The values $\mathcal{D}_{\mathrm{norm}} =(2.88 \pm 0.17) \times 10^{6}$ for $\Bs$
and $(2.77 \pm 0.16) \times 10^{6}$ for $\Bz$ are obtained
using Tables~\ref{tab:bplus} and \ref{tab:effratio} for each category,  
together with the combined uncertainty from Table~\ref{tab:systematics-2},
and including the $+4$\% correction to the \Bsmumu\ efficiency due to the
lifetime difference between $B_{s,\mathrm{H}}$ and $\Bs$. 

The combination of $\Bp$ branching fraction, hadronisation probabilities and
$\mathcal{D}_{\mathrm{norm}}$, i.e. the single-event sensitivity,
is equal to $(8.9\pm 1.0)\times 10^{-11}$ for 
\Bsmumu\ and $(2.21\pm0.15)\times 10^{-11}$ for \Bdmumu.

The values of the branching fractions that optimise the profile-likelihood within the
constraint of non-negative values are 
$\BR (\Bsmumu) =  0.9 \times 10^{-9}$ and $\BR (\Bdmumu) = 0$.   
That constraint is applied for all results discussed in this section
if not otherwise stated.

A Neyman construction~\cite{neyman} is used to determine the $68.3\%$ confidence interval
for \BR(\Bsmumu) with pseudo-MC experiments, obtaining:
\begin{equation}
\BR (\Bsmumu) =  \left( 0.9^{+1.1}_{-0.8} \right) \times 10^{-9}\, .\notag
\end{equation}
The uncertainties include both the statistical and systematic contributions. The two components
are separated by repeating the likelihood fit after setting all systematic uncertainties to zero.
The statistical uncertainty is dominant, with the systematic uncertainty equal to 
$\pm \,0.3\times 10^{-9}$.

The observed significance of the \Bsmumu\ signal is determined from pseudo-MC experiments, with
a hypothesis test based on the likelihood ratio 
$-\ln [L(\mathrm{no\!-\!signal})/L(\mathrm{max})]$~\cite{Cowan:2010js},
and is equal to 1.4 standard deviations. 
For this test, \BR(\Bdmumu) is left free to be determined in the fit.
The corresponding expected significance for the SM prediction is $3.1$ standard deviations.

Pseudo-MC experiments are also used to evaluate the compatibility of the observation with the SM prediction. 
A hypothesis test based on $-\ln [L(\mathrm{SM})/L(\mathrm{max})]$
is performed for the simultaneous fit to \BR(\Bsmumu) and \BR(\Bdmumu). 
The result is $p=0.048\pm0.002$, corresponding to $2.0$ standard deviations.

\begin{figure}[!t]
  \begin{center}
    \includegraphics[width=0.75\textwidth]{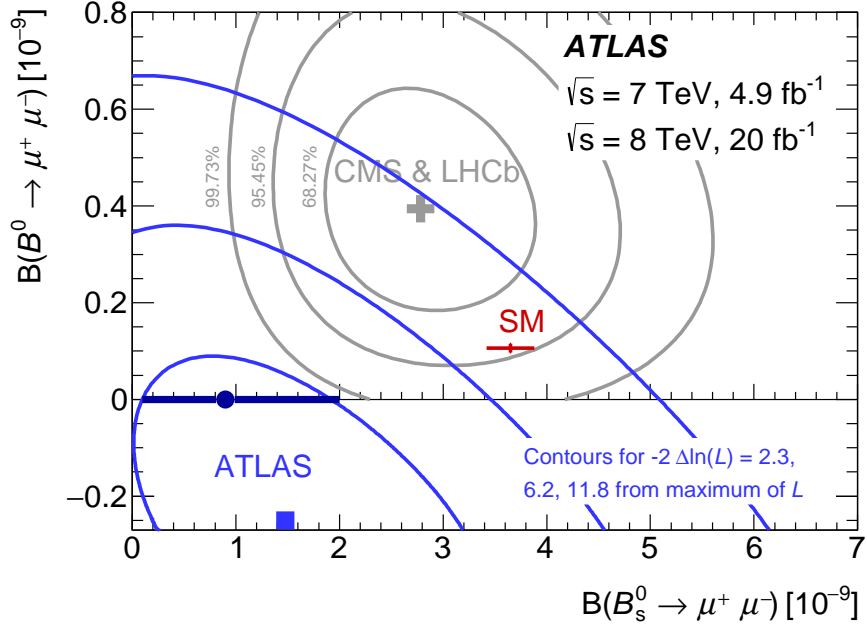}
    \caption
        {Contours in the plane $\BR (\Bsmumu), \BR (\Bdmumu)$ for intervals of
          $-2\, \Delta \ln(L)$ equal to $2.3$, $6.2$ and $11.8$ relative to the absolute maximum 
          of the likelihood, without imposing the constraint of non-negative branching fractions. 
          Also shown are the corresponding contours for the combined result of the CMS and LHCb 
          experiments, the SM prediction, and the maximum of the likelihood within the boundary of non-negative 
          branching fractions, with the error bars covering the 68.3\% confidence range for \BR(\Bsmumu).}
        \label{fig:like-contours-unconstrained}
  \end{center}
\end{figure}

Figure~\ref{fig:like-contours-unconstrained} shows the contours in the plane of \BR(\Bsmumu)
and \BR(\Bdmumu) drawn for values of $-2\, \Delta \ln(L)$ equal to $2.3$, $6.2$ and $11.8$,
relative to the maximum of the likelihood, allowing negative values of the branching fractions. 
The maximum within the physical boundary is shown with error bars 
indicating the $68.3$\% interval for the value of \BR(\Bsmumu).  
Also shown are the corresponding contours obtained in the combination of the results of the 
CMS and LHCb experiments, and the prediction based on the SM. 

Using the CL$_\mathrm{s}$ method~\cite{Read:2002hq} implemented with pseudo-MC experiments,
an upper limit is placed on the \Bsmumu\ branching fraction at the $95\%$ confidence level:
\begin{equation}
\BR(\Bsmumu) < 3.0\times 10^{-9}\: (95\% \: \mathrm{CL})\, . \notag
\end{equation}
The limit is obtained under the hypothesis of background only, with \BR(\Bdmumu) left free to be determined 
in the fit. The expected limit is $1.8^{+0.7}_{-0.4}  \times 10^{-9}$. 

An upper limit based on 
the CL$_\mathrm{s}$ method is also set on \BR(\Bdmumu). 
The expected limit obtained from pseudo-MC samples generated according to the observed amplitudes
of backgrounds and $\Bs$ signal is $\left( 5.7^{+2.1}_{-1.5} \right) \times 10^{-10}$ 
at a confidence level of $95\%$. The observed limit is:
\begin{equation}
\BR(\Bdmumu) < 4.2\times 10^{-10}\: (95\% \: \mathrm{CL})\, . \notag
\end{equation}
The observed upper limit is above the SM prediction and also covers
the central value of the combination of the measurements by CMS and LHCb~\cite{CMS:2014xfa}.
The expected significance for \BR(\Bdmumu) according to the SM prediction is equal to 
0.2 standard deviations.

%% file: BMuMuPaper_conclusions.tex
\section{Conclusions}
\label{sec:conclusion}

A study of the rare decays of \Bs and \Bz mesons into oppositely charged muon pairs is presented,
based on $25$~fb$^{-1}$ of $7$ \TeV\ and $8$ \TeV\ proton--proton collision data
collected by the ATLAS experiment in Run~1 of LHC.

For \Bz  an upper limit $\BR (\Bdmumu) <  4.2 \times 10^{-10}$ is placed at the 95\% confidence level,
based on the CL$_\mathrm{s}$ method. 
The limit is compatible with the predictions based on the SM
and with the combined result of the CMS and LHCb experiments.

For \Bs the result is $\BR (\Bsmumu) = \left( 0.9^{+1.1}_{-0.8} \right) \times 10^{-9}$, 
where the errors include both the statistical and systematic uncertainties. 
An upper limit $\BR (\Bsmumu) < 3.0 \times 10^{-9}$ at 95\% CL is placed,
lower than the SM prediction, and in better agreement with the measurement of CMS and LHCb.

A $p$-value of 4.8\% is found for the compatibility of the results with the SM prediction.

%% file: Acknowledgements.tex

We thank CERN for the very successful operation of the LHC, as well as the
support staff from our institutions without whom ATLAS could not be
operated efficiently.

We acknowledge the support of ANPCyT, Argentina; YerPhI, Armenia; ARC, Australia; BMWFW and FWF, Austria; ANAS, Azerbaijan; SSTC, Belarus; CNPq and FAPESP, Brazil; NSERC, NRC and CFI, Canada; CERN; CONICYT, Chile; CAS, MOST and NSFC, China; COLCIENCIAS, Colombia; MSMT CR, MPO CR and VSC CR, Czech Republic; DNRF and DNSRC, Denmark; IN2P3-CNRS, CEA-DSM/IRFU, France; GNSF, Georgia; BMBF, HGF, and MPG, Germany; GSRT, Greece; RGC, Hong Kong SAR, China; ISF, I-CORE and Benoziyo Center, Israel; INFN, Italy; MEXT and JSPS, Japan; CNRST, Morocco; FOM and NWO, Netherlands; RCN, Norway; MNiSW and NCN, Poland; FCT, Portugal; MNE/IFA, Romania; MES of Russia and NRC KI, Russian Federation; JINR; MESTD, Serbia; MSSR, Slovakia; ARRS and MIZ\v{S}, Slovenia; DST/NRF, South Africa; MINECO, Spain; SRC and Wallenberg Foundation, Sweden; SERI, SNSF and Cantons of Bern and Geneva, Switzerland; MOST, Taiwan; TAEK, Turkey; STFC, United Kingdom; DOE and NSF, United States of America. In addition, individual groups and members have received support from BCKDF, the Canada Council, CANARIE, CRC, Compute Canada, FQRNT, and the Ontario Innovation Trust, Canada; EPLANET, ERC, FP7, Horizon 2020 and Marie Sk{\l}odowska-Curie Actions, European Union; Investissements d'Avenir Labex and Idex, ANR, R{\'e}gion Auvergne and Fondation Partager le Savoir, France; DFG and AvH Foundation, Germany; Herakleitos, Thales and Aristeia programmes co-financed by EU-ESF and the Greek NSRF; BSF, GIF and Minerva, Israel; BRF, Norway; the Royal Society and Leverhulme Trust, United Kingdom.

The crucial computing support from all WLCG partners is acknowledged
gratefully, in particular from CERN and the ATLAS Tier-1 facilities at
TRIUMF (Canada), NDGF (Denmark, Norway, Sweden), CC-IN2P3 (France),
KIT/GridKA (Germany), INFN-CNAF (Italy), NL-T1 (Netherlands), PIC (Spain),
ASGC (Taiwan), RAL (UK) and BNL (USA) and in the Tier-2 facilities
worldwide.

%% file: atlas_authlist.tex
\begin{flushleft}
{\Large The ATLAS Collaboration}

\bigskip

M.~Aaboud$^\textrm{\scriptsize 136d}$,
G.~Aad$^\textrm{\scriptsize 87}$,
B.~Abbott$^\textrm{\scriptsize 114}$,
J.~Abdallah$^\textrm{\scriptsize 65}$,
O.~Abdinov$^\textrm{\scriptsize 12}$,
B.~Abeloos$^\textrm{\scriptsize 118}$,
R.~Aben$^\textrm{\scriptsize 108}$,
O.S.~AbouZeid$^\textrm{\scriptsize 138}$,
N.L.~Abraham$^\textrm{\scriptsize 150}$,
H.~Abramowicz$^\textrm{\scriptsize 154}$
% ...
\bigskip
\\
% ...
$^{12}$ Institute of Physics, Azerbaijan Academy of Sciences, Baku, Azerbaijan\\
% ...
$^{65}$ University of Iowa, Iowa City IA, United States of America\\
% ...
$^{87}$ CPPM, Aix-Marseille Universit{\'e} and CNRS/IN2P3, Marseille, France\\
% ...
$^{108}$ Nikhef National Institute for Subatomic Physics and University of Amsterdam, Amsterdam, Netherlands\\
% ...
$^{114}$ Homer L. Dodge Department of Physics and Astronomy, University of Oklahoma, Norman OK, United States of America\\
% ...
$^{118}$ LAL, Univ. Paris-Sud, CNRS/IN2P3, Universit{\'e} Paris-Saclay, Orsay, France\\
% ...
$^{136}$ $^{(a)}$ Facult{\'e} des Sciences Ain Chock, R{\'e}seau Universitaire de Physique des Hautes Energies - Universit{\'e} Hassan II, Casablanca; $^{(b)}$ Centre National de l'Energie des Sciences Techniques Nucleaires, Rabat; $^{(c)}$ Facult{\'e} des Sciences Semlalia, Universit{\'e} Cadi Ayyad, LPHEA-Marrakech; $^{(d)}$ Facult{\'e} des Sciences, Universit{\'e} Mohamed Premier and LPTPM, Oujda; $^{(e)}$ Facult{\'e} des sciences, Universit{\'e} Mohammed V, Rabat, Morocco\\
% ...
$^{138}$ Santa Cruz Institute for Particle Physics, University of California Santa Cruz, Santa Cruz CA, United States of America\\
% ...
$^{150}$ Department of Physics and Astronomy, University of Sussex, Brighton, United Kingdom\\
% ...
$^{154}$ Raymond and Beverly Sackler School of Physics and Astronomy, Tel Aviv University, Tel Aviv, Israel\\
% ...
$^{a}$ Also at Department of Physics, King's College London, London, United Kingdom\\
$^{b}$ Also at Institute of Physics, Azerbaijan Academy of Sciences, Baku, Azerbaijan\\
$^{c}$ Also at Novosibirsk State University, Novosibirsk, Russia\\
$^{d}$ Also at TRIUMF, Vancouver BC, Canada\\
$^{e}$ Also at Department of Physics {\&} Astronomy, University of Louisville, Louisville, KY, United States of America\\
% ...
\end{flushleft}
